\documentclass[10pt,twoside]{amundi_article}

\usepackage[table]{xcolor}
\usepackage[latin9]{inputenc}
\usepackage{amssymb}
\usepackage{amsfonts}
\usepackage{amsmath}
\usepackage{amsthm}
\usepackage{fancybox}
\usepackage{fancyhdr}
\usepackage{graphicx}
\usepackage{subcaption}
\usepackage[hyphens]{url}
\usepackage{arydshln}
\usepackage{multirow}
\usepackage{pdflscape}
\usepackage{tikz}
\usepackage{comment}
\usepackage{nicefrac}
\usepackage{subcaption}
\usepackage{dsfont}
\usepackage{algorithmic}
\usepackage{algorithm}

\newlength{\figurewidth}
\newlength{\figureheight}
\def\figureskip{\vskip 10pt plus 2pt minus 2pt\relax}

\newtheorem{remark}{Remark}
\def\limfunc#1{\mathop{\rm #1}}
\def\func#1{\mathop{\rm #1}}

\newcommand{\TsV}{\hspace{5pt}}
\newcommand{\TsVIII}{\hspace{8pt}}

\newcommand*\LogL{\ensuremath{\boldsymbol\ell}}

\begin{document}

\setcounter{page}{1}

\title{\textbf{\color{amundi_blue}Financial Applications of Gaussian Processes and Bayesian Optimization}%
\footnote{We would like to thank Elisa Baku and Thibault Bourgeron for their helpful comments.}}
\author{
{\color{amundi_dark_blue} Joan Gonzalvez} \\
Quantitative Research \\
Amundi Asset Management, Paris \\
\texttt{joan.gonzalvez@amundi.com} \and
{\color{amundi_dark_blue} Edmond Lezmi} \\
Quantitative Research \\
Amundi Asset Management, Paris \\
\texttt{edmond.lezmi@amundi.com} \and
{\color{amundi_dark_blue} Thierry Roncalli} \\
Quantitative Research \\
Amundi Asset Management, Paris \\
\texttt{thierry.roncalli@amundi.com} \and
{\color{amundi_dark_blue} Jiali Xu} \\
Quantitative Research \\
Amundi Asset Management, Paris \\
\texttt{jiali.xu@amundi.com}}

\date{\color{amundi_dark_blue}February 2019}

\maketitle

\begin{abstract}
In the last five years, the financial industry has been impacted by the
emergence of digitalization and machine learning. In this article, we explore
two methods that have undergone rapid development in  recent years:
Gaussian processes and Bayesian optimization. Gaussian processes can be seen
as a generalization of Gaussian random vectors and are associated with the
development of kernel methods. Bayesian optimization is an approach for
performing derivative-free global optimization in a small dimension, and uses
Gaussian processes to locate the global maximum of a black-box function. The
first part of the article reviews these two tools and shows how they are
connected. In particular, we focus on the Gaussian process regression, which
is the core of Bayesian machine learning, and the issue of hyperparameter
selection. The second part is dedicated to two financial applications. We
first consider the modeling of the term structure of interest rates. More
precisely, we test the fitting method and compare the GP prediction and the
random walk model. The second application is the construction of
trend-following strategies, in particular the online estimation of trend and
covariance windows.
\end{abstract}

\noindent \textbf{Keywords:} Gaussian process, Bayesian optimization, machine
learning, kernel function, hyperparameter selection, regularization,
time-series prediction, asset allocation, portfolio optimization,
trend-following strategy, moving-average estimator, ADMM, Cholesky trick.\medskip

\noindent \textbf{JEL classification:} C61, C63, G11.

\clearpage

\section{Introduction}

This article explores the use of Gaussian processes and Bayesian optimization
in finance. These two tools have been successful in the machine learning
community. In recent years, machine learning algorithms have been applied
in risk management, asset management, option trading and market making. Despite
the skepticism about earlier implementations, we must today recognize that
machine learning is changing the world of finance. Banks, asset managers, hedge
funds and robo-advisors have invested a lot of money in such technologies, and
all the reports agree that it is just the beginning (McKinsey, 2015; OECD,
2017; Oliver Wyman, 2018). Even supervisory bodies are closely monitoring this
development and its impact on the financial industry (FSB, 2017). Certainly,
the most impressive indicator is the evolution of the financial job market
(BCG, 2018). Today, applicants in the quant finance must have a certification
in machine learning or at least knowledge of this technology and
experience in the Python programming language.\smallskip

In our two previous works, we focus on asset allocation and portfolio construction.
In Bourgeron \textsl{et al.} (2018), we discuss how to design a comprehensive
and automated portfolio optimization model for robo-advisors. In Richard and
Roncalli (2019), we extend this approach when we consider risk budgeting
portfolios in place of mean-variance portfolios. This third paper continues the
\textquoteleft \textit{tour d'horizon}\textquoteright\ of machine learning
techniques that can be useful for asset management challenges. However, we are
significantly changing the direction since we move away from portfolio
optimization and we are interested here in estimation and forecasting
problems.\smallskip

A Gaussian process (GP) is generalization of a Gaussian random vector, and can
be seen as a stochastic process on general continuous functions. This can be
done because it replaces the traditional covariance matrix by a kernel
function, and benefits from the power of kernel methods. The core of this
approach is the computation of the conditional distribution. In a Bayesian
framework, this is equivalent to computing the posterior distribution from the
prior distribution. Since linear regression is the solution of the conditional
expectation problem when the random variables are Gaussian, it is then
straightforward to define Gaussian process regression, which is a powerful
semi-parametric machine learning model (Rasmussen and Williams, 2006) that has
been successfully used in geostatistics (Cressie, 1993), multi-task learning
(Alvarez \textsl{et al.}, 2012) or robotics and reinforcement learning
(Deisenroth \textsl{et al.}, 2015). Bayesian optimization is an approach used to
solve black-box optimization problems (Bochu \textsl{et al.}, 2009; Frazier,
2018) where the objective function is not explicitly known and costly to
evaluate. Without access to the gradient vector of the objective function, usual
quasi-Newton or gradient-descent methods are unusable. Bayesian optimization
models the unknown function as a random Gaussian process surrogate and replaces
the intractable original problem by a sequence of simpler optimization
problems. In this case, GPs appear as a tool in Bayesian optimization.
Generally, a financial model depends on some external parameters that have to
be fixed before running the model. Bayesian optimization mainly concerns the
estimation of these external parameters, which are called hyperparameters.
Examples are the length of a moving-average estimator, the risk aversion of the
investor or the window of a covariance matrix of asset returns.\smallskip

This paper is organized as follows. Section Two reviews the mathematics of
Gaussian processes and Bayesian optimization. In particular, we present the
technique of Gaussian process regression and discuss the issue of hyperparameter
selection. In Section Three, we use Gaussian processes in order to fit the term
structure of the interest rates, and show how they can be used for forecasting
the yield curve. The second application of Section Three concerns the online
estimation of the hyperparameters of the trend-following strategy. Finally,
Section Four offers some concluding remarks.

\section{A primer on Gaussian processes and Bayesian optimization}

In this section, we define the main concepts and techniques used in machine
learning with Gaussian processes. In regression and classification problems,
they are used for interpolation, extrapolation and pattern discovery for which
the choice of the kernel function is central. Contrary to many supervised
learning algorithms, GPs have the distinctive property of estimating the
variance of the prediction or the confidence region for a test sample. This
feature is useful for global optimization and helps to improve the objective
function, because validation samples can be evaluated according to the
confidence of the model.\smallskip

Bayesian optimization is a statistical approach used to solve black-box optimization
problems, where the objective function is not explicitly known or costly to
evaluate. Since the gradient of the objective function is difficult to evaluate
or unknown, descent methods are unusable. In this case, Bayesian optimization
replaces the unknown objective function by a random Gaussian process and the
intractable original problem by a sequence of simpler optimization problems.
This explains why Gaussian processes and Bayesian optimization are closely
related.

\subsection{Gaussian processes}

\subsubsection{Definition}

Let $\mathcal{X}$ be a set\footnote{The set $\mathcal{X}$ of inputs can be
multi-dimensional (linear regression modeling), time-dimensional (time series
forecasting), etc.} in $\mathbb{R}^{d}$. A Gaussian process is a collection
$\left\{ f\left( x\right) ,\ x\in \mathcal{X}\right\} $ such that for any $n\in
\mathbb{N}$ and $x_{1},\dots ,x_{n}\in \mathcal{X}$, the random vector $\left(
f\left( x_{1}\right) ,\dots ,f\left( x_{n}\right) \right) $ has a joint
multivariate Gaussian distribution (Rasmussen and Williams, 2006). Therefore,
we can characterize the GP by its mean function:
\begin{equation*}
m\left( x\right) =\mathbb{E}\left[ f\left( x\right) \right]
\end{equation*}%
and its covariance function:%
\begin{eqnarray*}
\mathcal{K}\left( x,x^{\prime }\right)  &=&\limfunc{cov}\left( f\left(
x\right) ,f\left( x^{\prime }\right) \right)  \\
&=&\mathbb{E}\left[ \left( f\left( x\right) -m(x)\right) \left( f\left(
x^{\prime }\right) -m\left( x^{\prime }\right) \right) \right]
\end{eqnarray*}%
The covariance function $\mathcal{K}\left( x,x^{\prime }\right) $ is central in
the analysis of Gaussian processes, and is called the \textquoteleft
\textit{kernel}\textquoteright\ function. In machine learning, the most popular
kernel function is the squared exponential kernel\footnote{It is also known as
the radial basis function (RBF) kernel or the Gaussian kernel.}, which is given
by:
\begin{equation}
\mathcal{K}_{\mathrm{SE}}\left( x,x^{\prime }\right) =\exp \left( -
\frac{1}{2}\left\Vert x-x^{\prime }\right\Vert _{2}^{2}\right)   \label{gp1}
\end{equation}
for $x\in \mathbb{R}^{d}$ and $x^{\prime }\in \mathbb{R}^{d}$.

\begin{remark}
In what follows, we assume that $m\left( x\right) =\mathbf{0}$ without loss of
generality.
\end{remark}

\subsubsection{Gaussian process regression}

Given a training set $\left\{ \left( x_{i},y_{i}\right) \right\} _{i=1}^{n}$ of
features, the goal of Gaussian process regression (GPR) is to forecast $f\left(
x^{\star }\right) $ for some new inputs $x^{\star }\in \mathbb{R}^{n^{\star
}\times d}$. For that, we adopt the Bayesian framework, and we compute the
posterior distribution of the GP conditionally on $x=\left( x_{1},\dots
,x_{n}\right) \in \mathbb{R}^{n\times d}$.

\paragraph{The noise-free case}

Let us first assume that the observations are noise-less, meaning that
$y=f\left( x\right) =\left( f\left( x_{1}\right) ,\dots ,f\left( x_{n}\right)
\right) $ where $f$ is the GP. Let $x^{\star }$ be $n^{\star }$ new inputs and
$\hat{y}^{\star }=\mathbb{E}\left[ f\left( x^{\star } \mid x,y\right)\right] $
be the conditional prediction. Then we have:
\begin{equation}
f\left( x,x^{\star }\right) \sim \mathcal{N}\left( \mathbf{0}_{n+n^{\star
}},\left(
\begin{array}{cc}
\mathcal{K}\left( x,x\right)  & \mathcal{K}\left( x,x^{\star }\right)  \\
\mathcal{K}\left( x^{\star },x\right)  & \mathcal{K}\left( x^{\star
},x^{\star }\right)
\end{array}%
\right) \right)   \label{gp:prior1}
\end{equation}%
where $\mathcal{K}\left( x^{\star },x\right) $ is the $n^{\star }\times n$
matrix\footnote{We must not confuse the kernel function $\mathcal{K}\left(
x,x^{\prime }\right) $ where $x\in \mathbb{R}^{d}$ and $x^{\prime }\in
\mathbb{R}^{d}$ that returns a scalar, and the kernel matrix $\mathcal{K}\left(
x^{\star },x\right) $ where $x\in \mathbb{R}^{n\times d}$ and $x^{\star }\in
\mathbb{R}^{n^{\star }\times d}$ that returns a $n^{\star }\times n$ matrix.}
with entries $\mathcal{K}_{i,j}\left( x^{\star },x\right) =\mathcal{K} \left(
x_{i}^{\star },x_{j}\right) $. Using Appendix
\ref{appendix:section-conditional-expectation} on page
\pageref{appendix:section-conditional-expectation}, it follows that the random
vector $y^{\star }\mid x,y=f\left( x^{\star }\mid x,y\right) $ is also
Gaussian:
\begin{equation*}
f\left( x^{\star }\mid x,y\right) \sim \mathcal{N}\left( m\left( x^{\star
}\mid x,y\right) ,\mathcal{K}\left( x^{\star },x^{\star }\mid x,y\right)
\right)
\end{equation*}
where $m\left( x^{\star }\mid x,y\right) $ is the mean vector of the
posterior distribution\footnote{%
We reiterate that $m\left( x\right) =\mathbf{0}_{n}$ and $m\left( x^{\star
}\right) =\mathbf{0}_{n^{\star }}$.}:%
\begin{equation*}
m\left( x^{\star }\mid x,y\right) =\mathcal{K}\left( x^{\star },x\right) \mathcal{K}\left( x,x\right) ^{-1}y
\end{equation*}%
and the covariance matrix $\mathcal{K}\left( x^{\star },x^{\star }\mid
x,y\right) $
is the \textit{Schur's complement} of the prior:%
\begin{equation*}
\mathcal{K}\left( x^{\star },x^{\star }\mid x,y\right) =\mathcal{K}\left(
x^{\star },x^{\star }\right) -\mathcal{K}\left( x^{\star },x\right) \mathcal{%
K}\left( x,x\right) ^{-1}\mathcal{K}\left( x,x^{\star }\right)
\end{equation*}%
We deduce that the prediction is the conditional expectation:%
\begin{equation*}
\hat{y}^{\star }=m\left( x^{\star }\mid x,y\right)
\end{equation*}%
We notice that computing the posterior distribution requires us to invert the $%
n\times n$ matrix $\mathcal{K}\left( x,x\right) $. Since it is a covariance
matrix, it is a symmetric positive semi-definite matrix and the Cholesky
decomposition can be applied leading to $O\left( n^{3}\right) $ operations.

\begin{remark}
In order to reduce the notation complexity, we introduce the hat notation for
writing conditional quantities. We have $\hat{f}\left( x^{\star }\right)
=f\left( x^{\star }\mid x,y\right) $, $\hat{m}\left( x^{\star }\right) =m\left(
x^{\star }\mid x,y\right) $ and $\mathcal{\hat{K}}\left( x^{\star },x^{\star
}\right) =\mathcal{K}\left( x^{\star },x^{\star }\mid x,y\right) $.
\end{remark}

\paragraph{Gaussian noise}

In order to take into account noise in the data, we assume that $y=f\left(
x\right) +\varepsilon $ where $\varepsilon \sim \mathcal{N}\left( \mathbf{0}%
_{n},\sigma_{\varepsilon}^{2}I_{n}\right) $. In this case, Equation
(\ref{gp:prior1})
becomes:%
\begin{equation}
f\left( x,x^{\star }\right) \sim \mathcal{N}\left( \mathbf{0}_{n+n^{\star
}},\left(
\begin{array}{cc}
\mathcal{K}\left( x,x\right) +\sigma_{\varepsilon} ^{2}I_{n} & \mathcal{K}\left(
x,x^{\star }\right)  \\
\mathcal{K}\left( x^{\star },x\right)  & \mathcal{K}\left( x^{\star
},x^{\star }\right)
\end{array}%
\right) \right)   \label{gp:prior2}
\end{equation}%
Again, the posterior distribution is Gaussian and we have:%
\begin{equation*}
\hat{f}\left( x^{\star }\right) \sim \mathcal{N}\left( \hat{m}\left(
x^{\star }\right) ,\mathcal{\hat{K}}\left( x^{\star },x^{\star }\right)
\right)
\end{equation*}%
where:%
\begin{equation*}
\hat{m}\left( x^{\star }\right) =\mathcal{K}\left( x^{\star },x\right)
\left( \mathcal{K}\left( x,x\right) +\sigma_{\varepsilon} ^{2}I_{n}\right) ^{-1}y
\end{equation*}%
and:%
\begin{equation*}
\mathcal{\hat{K}}\left( x^{\star },x^{\star }\right) =\mathcal{K}\left(
x^{\star },x^{\star }\right) -\mathcal{K}\left( x^{\star },x\right) \left(
\mathcal{K}\left( x,x\right) +\sigma_{\varepsilon} ^{2}I_{n}\right) ^{-1}\mathcal{K}\left(
x,x^{\star }\right)
\end{equation*}

\paragraph{Scalability issues}

For large datasets, inverting the kernel matrix $\mathcal{K}\left( x,x\right) $
leads to a $O\left( n^{3}\right) $ complexity and may be prohibitive.
Therefore, several methods have been proposed to adapt naive GPR to such
problems (Qui\~nonero-Candela and Rasmussen, 2015; Qui\~nonero-Candela
\textsl{et al.}, 2007). For instance, the subsets of regressors (SoR) algorithm
uses a low-rank approximation of the matrix $\mathcal{K}\left( x,x\right) $. If
we select $m<n$ samples $x_{m}$ from the training set $x$, the approximation of
$\mathcal{K}\left( x,x^{\prime }\right) $ is given by\footnote{$x_{m}$ are
called the \textquoteleft \textit{inducing points}\textquoteright.}:
\begin{equation*}
\mathcal{K}\left( x,x^{\prime }\right) \approx \mathcal{K}\left(
x,x_{m}\right) \mathcal{K}\left( x_{m},x_{m}\right) ^{-1}\mathcal{K}\left(
x_{m},x^{\prime }\right)
\end{equation*}%
In Appendix \ref{appendix:section-sor-approximation} on page \pageref{appendix:section-sor-approximation}, we show that:%
\begin{equation*}
\hat{m}\left( x^{\star }\right) \approx \mathcal{K}\left( x^{\star
},x_{m}\right) \mathcal{\tilde{K}}\left( x_{m},x_{m}\right) ^{-1}\mathcal{K}%
\left( x_{m},x\right) y
\end{equation*}%
and:%
\begin{equation*}
\mathcal{\hat{K}}\left( x^{\star },x^{\star }\right) \approx \sigma_{\varepsilon} ^{2}%
\mathcal{K}\left( x^{\star },x_{m}\right) \mathcal{\tilde{K}}\left(
x_{m},x_{m}\right) ^{-1}\mathcal{K}\left( x_{m},x^{\star }\right)
\end{equation*}%
where:%
\begin{equation*}
\mathcal{\tilde{K}}\left( x_{m},x_{m}\right) =\mathcal{K}\left(
x_{m},x\right) \mathcal{K}\left( x,x_{m}\right) +\sigma_{\varepsilon} ^{2}\mathcal{K}%
\left( x_{m},x_{m}\right)
\end{equation*}%
Gaussian process regression can then be done by inverting the $m\times m$
matrix $\mathcal{\tilde{K}}\left( x_{m},x_{m}\right) $ instead of the $%
n\times n$ matrix $\mathcal{K}\left( x,x\right) +\sigma_{\varepsilon}
^{2}I_{n}$.

\begin{remark}
Other methods to reduce the computational cost of GPR include Bayesian
committee machines (BCM) introduced by Tresp (2000). The underlying idea is to
train several Gaussian processes (or other kernel machines) on subsets of data
and mix them according to their prediction confidence. This kind of ensemble
methods can be particularly useful for large-scale problems and have been
designed for parallel computation (Deisenroth and Ng, 2015; Liu \textsl{et
al.}, 2018).
\end{remark}

\subsubsection{Covariance functions}

Gaussian processes can be seen as probability distributions over functions.
Therefore, the covariance function of the GP determines the properties of the
function $f\left( x\right) $. For instance, $\mathcal{K}\left( x,x^{\prime
}\right) =\min \left( x,x^{\prime }\right) $ is the covariance function of the
Brownian motion, thus samples from a GP with this kernel function will be
nowhere differentiable. Regularity, periodicity and monotonicity of the samples
can all be controlled by choosing the appropriate kernel. Moreover, operations
on kernels allow us to extract more complex patterns and structures in the data
(Duvenaud \textsl{et al}., 2013). Kernels can be summed, multiplied and
convoluted, yielding another valid covariance function (Bishop, 2006). In what
follows, we introduce the most used covariance kernels and their properties.

\paragraph{Usual covariance kernels}

A simple covariance function is the linear kernel, which is given by:
\begin{equation*}
\mathcal{K}_{\mathrm{Linear}}\left( x,x^{\prime }\right) =x^{\top }x^{\prime}
\end{equation*}%
GP regression using the linear kernel is equivalent to Bayesian linear
regression and multiplying it by itself several times yields Bayesian
polynomial regression.\smallskip

One of the most used covariance functions is the SE kernel mentioned above,
which can be generalized in the following way:
\begin{equation*}
\mathcal{K}_{\mathrm{SE}}\left( x,x^{\prime }\right) =\sigma ^{2}\exp \left(
-\frac{1}{2}\left( x-x^{\prime }\right) ^{\top }\Sigma \left( x-x^{\prime
}\right) \right)
\end{equation*}%
where $\Sigma $ is the $d\times d$ matrix that parameterizes the length scales
of the inputs. Taking $\Sigma =\func{diag}\left( \ell _{1}^{2},\dots ,\ell
_{d}^{2}\right) $ allows us to scale each dimension of the inputs separately.
Setting $\ell _{j}=0$ will eliminate the $j^{\mathrm{th}}$ dimension of the
input, which can be useful when constructing complex kernels. This kernel is
sometimes called the automatic relevance determination (ARD) kernel since it
can be used to discover relevant dimensions of the inputs when optimizing the
hyperparameters $\left( \ell _{1},\dots ,\ell _{d}\right) $. Sample functions
with this covariance kernel have infinitely many derivatives.\smallskip

Adding together SE kernels with different length scales gives the rational
quadratic (RQ) kernel. Let us consider the Gamma distribution parameterized by
shape $\alpha $ and rate $\beta $, whose density function is equal to:
\begin{equation*}
g_{\alpha ,\beta }\left( x\right) =\frac{\beta ^{\alpha }}{\Gamma \left(
\alpha \right) }x^{\alpha -1}e^{-\beta x}
\end{equation*}%
If we consider a Bayesian prior on the inverse squared length scale $\tau $
using this Gamma distribution, we obtain:
\begin{equation*}
\mathcal{K}_{\mathrm{SE}}\left( x,x^{\prime }\mid \tau \right) =\sigma
^{2}e^{-\frac{1}{2}\tau r^{2}}
\end{equation*}%
where $r=\left\Vert x-x^{\prime }\right\Vert _{2}$. Then, we have:%
\begin{eqnarray*}
\mathcal{K}_{\mathrm{RQ}}\left( x,x^{\prime }\right)  &=&\int_{0}^{+\infty }%
\mathcal{K}_{\mathrm{SE}}\left( x,x^{\prime }\mid \tau \right) g_{\alpha
,\beta }\left( \tau \right) \,\mathrm{d}\tau  \\
&=&\sigma ^{2}\frac{\beta ^{\alpha }}{\Gamma \left( \alpha \right) }%
\int_{0}^{+\infty }\tau ^{\alpha -1}e^{-\left( \beta +\frac{1}{2}%
r^{2}\right) \tau }\,\mathrm{d}\tau  \\
&=&\sigma ^{2}\frac{\beta ^{\alpha }}{\Gamma \left( \alpha \right) }\left[
-\left( \beta +\frac{1}{2}r^{2}\right) ^{-\alpha }\Gamma \left( \alpha
,\left( \beta +\frac{1}{2}r^{2}\right) \tau \right) \right] _{0}^{\infty } \\
&\propto &\frac{1}{\left( \beta +\frac{1}{2}r^{2}\right) ^{\alpha }} \\
&\propto &\left( 1+\frac{\left\Vert x-x^{\prime }\right\Vert _{2}^{2}}{%
2\alpha \ell ^{2}}\right) ^{-\alpha }
\end{eqnarray*}%
where $\Gamma \left( \alpha ,x\right) =\int_{x}^{+\infty }x^{\alpha
-1}e^{-x}\,\mathrm{d}x$ is the upper incomplete gamma function and $\beta =\ell
^{2}\alpha $ for a given $\ell $. We deduce that the RQ kernel is
isotropic, because it only depends on the Euclidean norm $r=\left\Vert
x-x^{\prime }\right\Vert _{2}$.\smallskip

Another class of popular kernels is the Mat\'ern family given by:
\begin{equation*}
\mathcal{K}_{\mathrm{Matern}}\left( x,x^{\prime }\right) =\frac{2^{1-\nu }}{%
\Gamma \left( \nu \right) }\left( \frac{\sqrt{2\nu }\left\Vert x-x^{\prime
}\right\Vert _{2}}{\ell }\right) ^{\nu }K_{\nu }\left( \frac{\sqrt{2\nu }%
\left\Vert x-x^{\prime }\right\Vert _{2}}{\ell }\right)
\end{equation*}%
where $\nu $ and $\ell $ are two positive parameters and $K_{\nu }$ is the
modified Bessel function of the second kind. The normalization constant is such
that $\lim_{\nu \rightarrow \infty }\mathcal{K}_{\mathrm{Matern}}\left(
x,x^{\prime }\right) =\mathcal{K}_{\mathrm{SE}}\left( x,x^{\prime }\right) $.
This kernel appears quite complex but attractive, because its expression can be
simplified when $\nu $ is a half-integer (Rasmussen and Williams, 2006). For
instance, we have:
\begin{eqnarray*}
\mathcal{K}_{\mathrm{Matern32}}\left( x,x^{\prime }\right)  &=&\mathcal{K}_{%
\mathrm{Matern}}\left( x,x^{\prime };\frac{3}{2}\right)  \\
&=&\left( 1+\frac{\sqrt{3}r}{\ell }\right) \exp \left( -\frac{\sqrt{3}r}{%
\ell }\right)
\end{eqnarray*}%
and:%
\begin{eqnarray*}
\mathcal{K}_{\mathrm{Matern52}}\left( x,x^{\prime }\right)  &=&\mathcal{K}_{%
\mathrm{Matern}}\left( x,x^{\prime };\frac{5}{2}\right)  \\
&=&\left( 1+\frac{\sqrt{5}r}{\ell }+\frac{5r^{2}}{3\ell ^{2}}\right) \exp
\left( -\frac{\sqrt{5}r}{\ell }\right)
\end{eqnarray*}%
\smallskip

\begin{figure}[tbph]
\caption{Sample path of Gaussian processes with different kernel functions}
\label{fig:gpr1}
\begin{subfigure}[b]{0.49\linewidth}
\includegraphics[width=\linewidth]{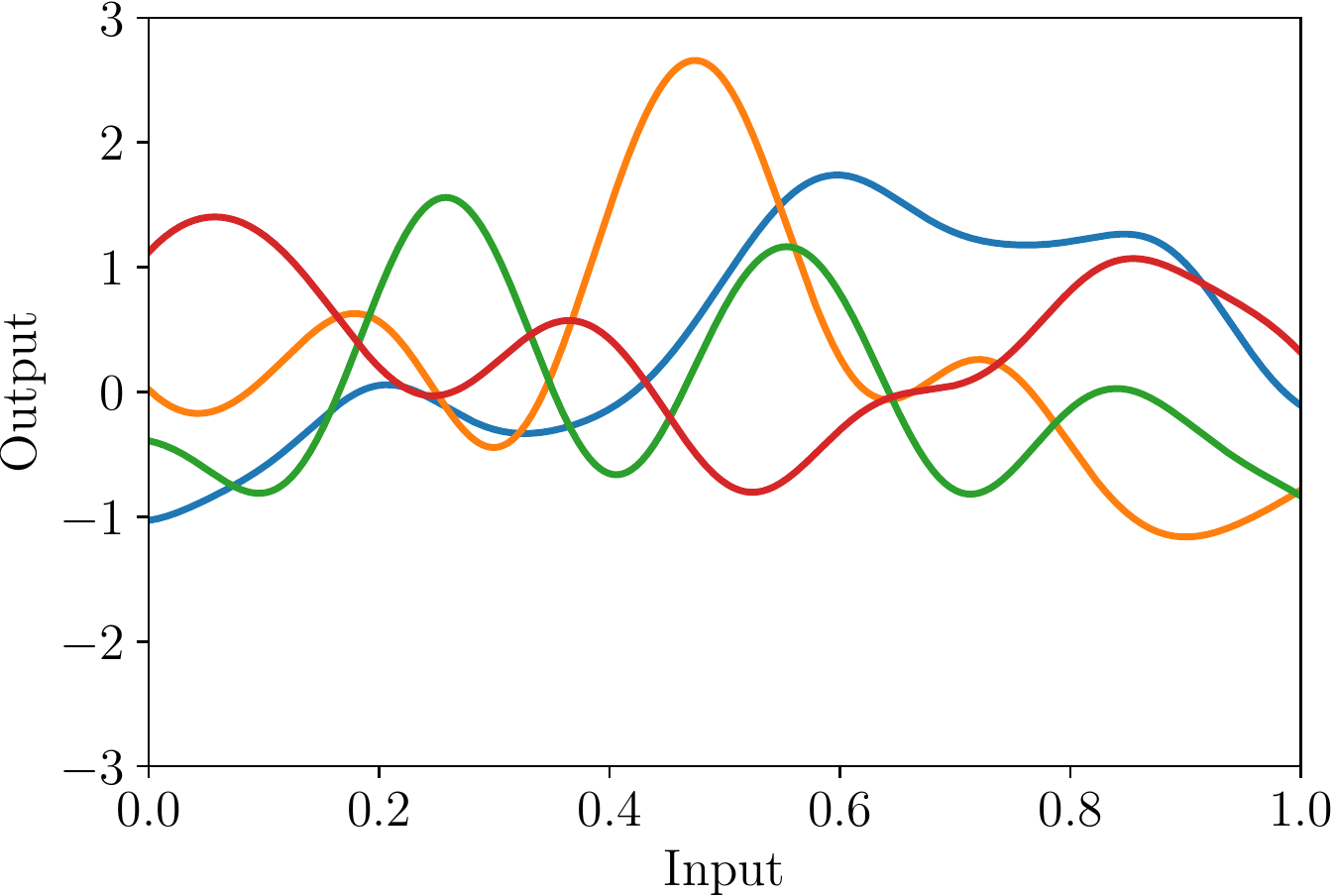}
\caption{$\mathcal{K}_{\mathrm{SE}} $}
\end{subfigure}
\begin{subfigure}[b]{0.49\linewidth}
\centering
\includegraphics[width=\linewidth]{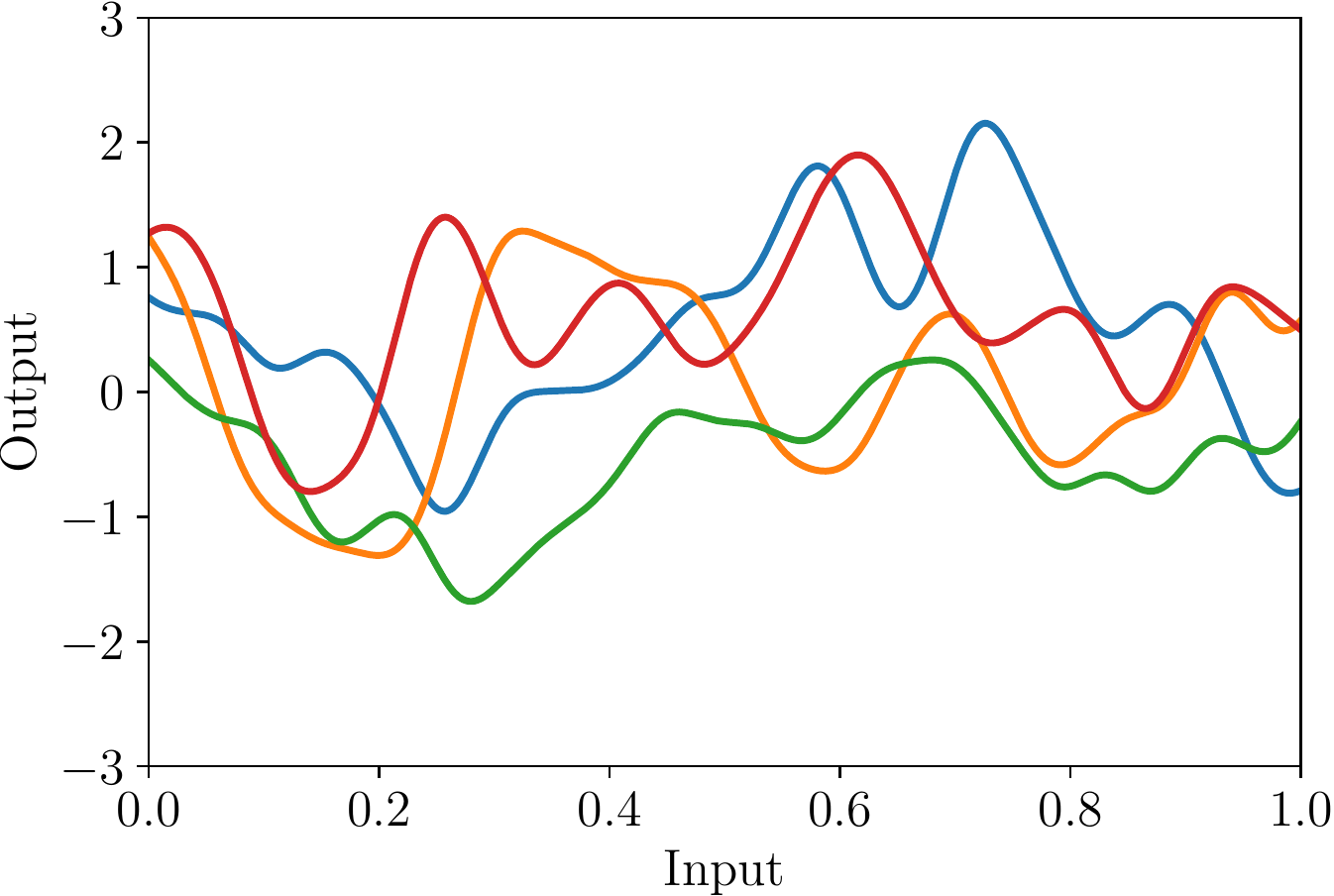}
\caption{$\mathcal{K}_{\mathrm{RQ}} $}
\end{subfigure}
\begin{subfigure}[b]{0.49\linewidth}
\centering
\includegraphics[width=\linewidth]{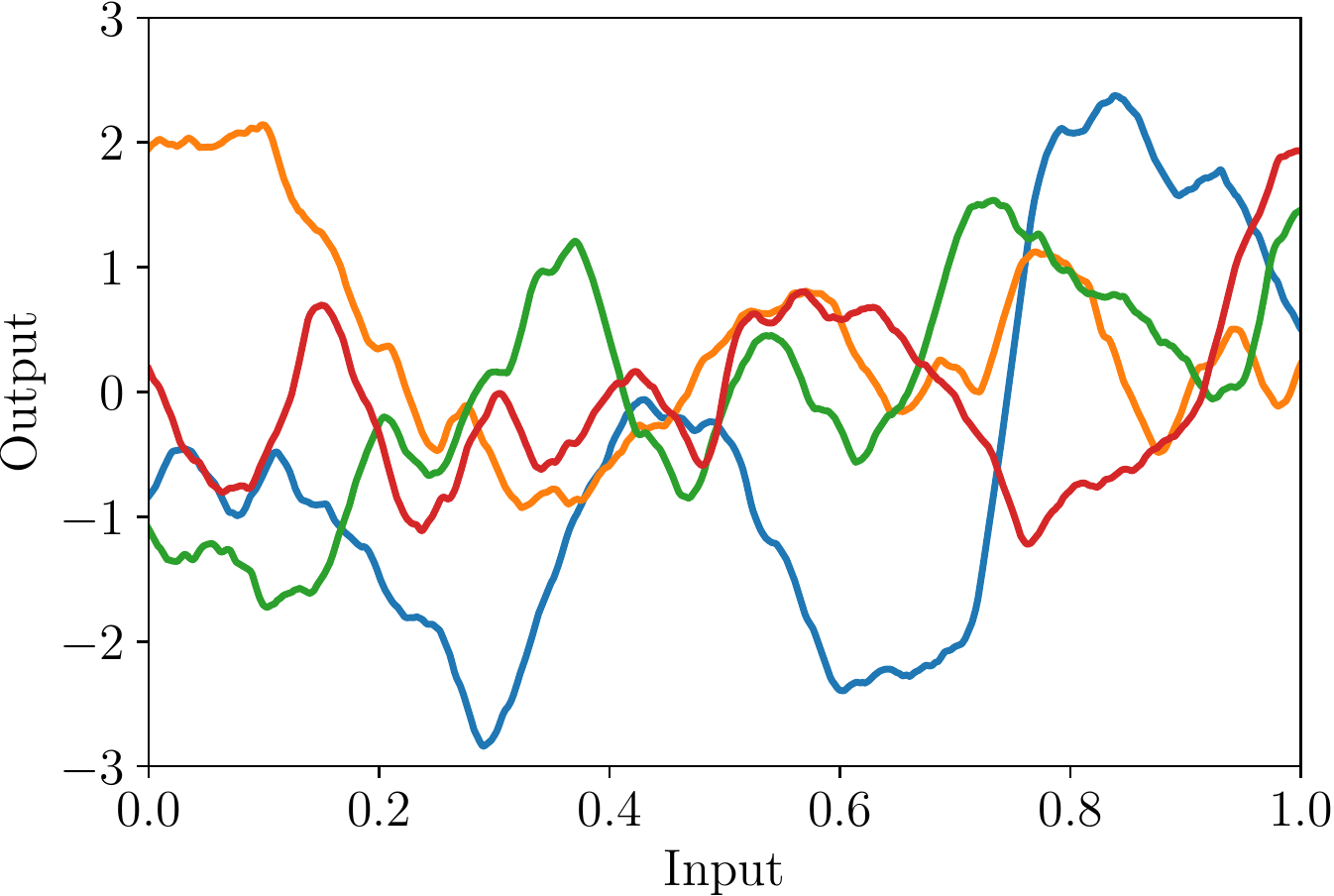}
\caption{$\mathcal{K}_{\mathrm{Matern32}} $}
\end{subfigure}
\begin{subfigure}[b]{0.49\linewidth}
\centering
\includegraphics[width=\linewidth]{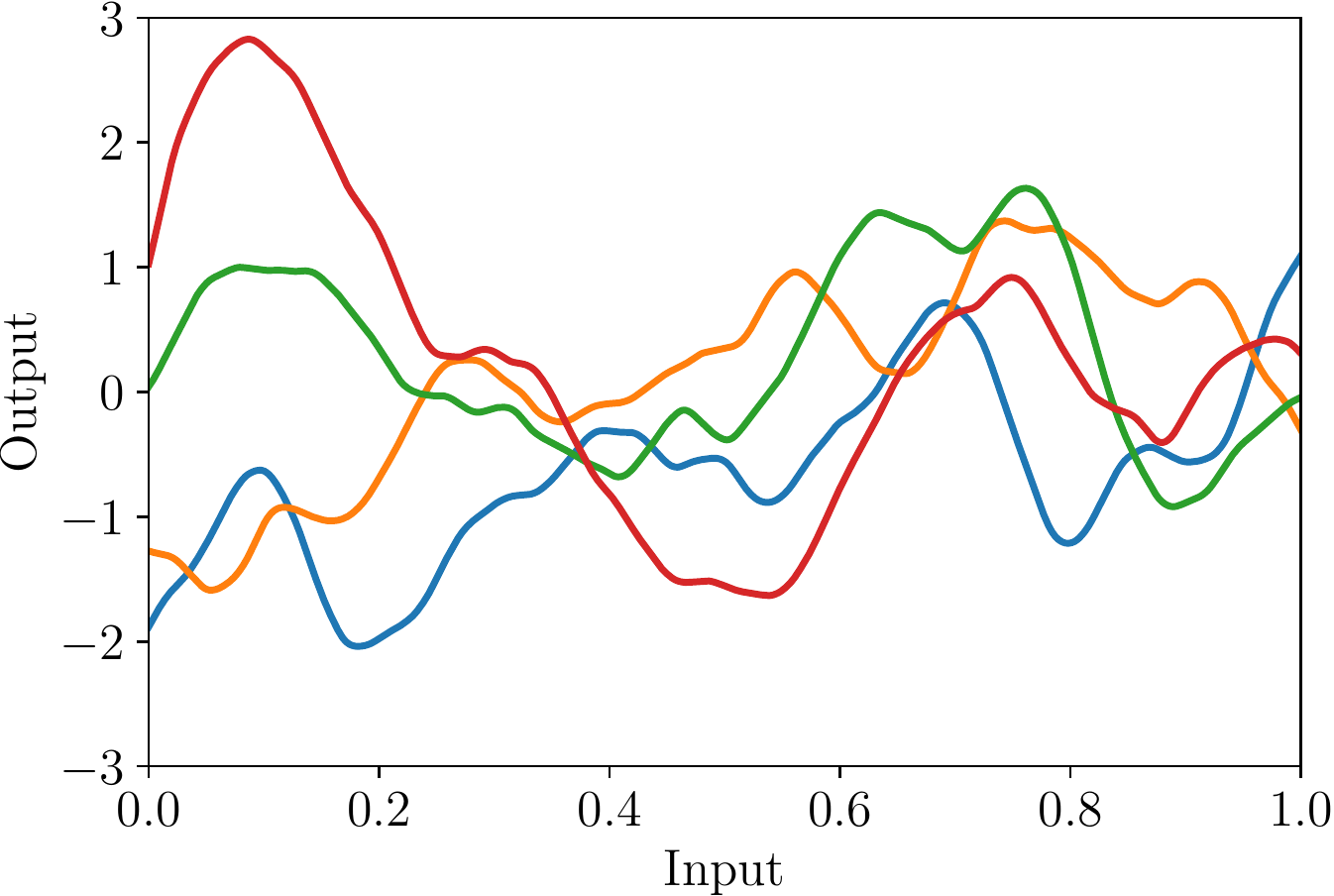}
\caption{$\mathcal{K}_{\mathrm{Matern52}} $}
\end{subfigure}
\end{figure}

Figure \ref{fig:gpr1} shows four sample paths of a one-dimensional Gaussian
process with different covariance functions. They are generated from a
multivariate Gaussian random vector and the Cholesky factorization of the
covariance matrix\footnote{We have $X=PU$ where $U\sim \mathcal{N}\left(
\mathbf{0}_{n},I_{n}\right) $, $P$ is the Cholesky decomposition of
$\mathcal{K}\left( x,x\right) $ such that $PP^{\intercal }=\mathcal{K}\left(
x,x\right) $, $\mathcal{K}\left( x,x\right) $ is the $n\times n$ covariance
matrix and $x$ is the uniform vector on the range $\left[ -1,1\right] $.}.

\begin{figure}[tbph]
\centering
\caption{Posterior distribution of $\func{sinc}\left(2x\right)$ ($\mathcal{K}_{\mathrm{SE}}$ kernel)}
\label{fig:gpr2a}
\figureskip
\includegraphics[width = \figurewidth, height = \figureheight]{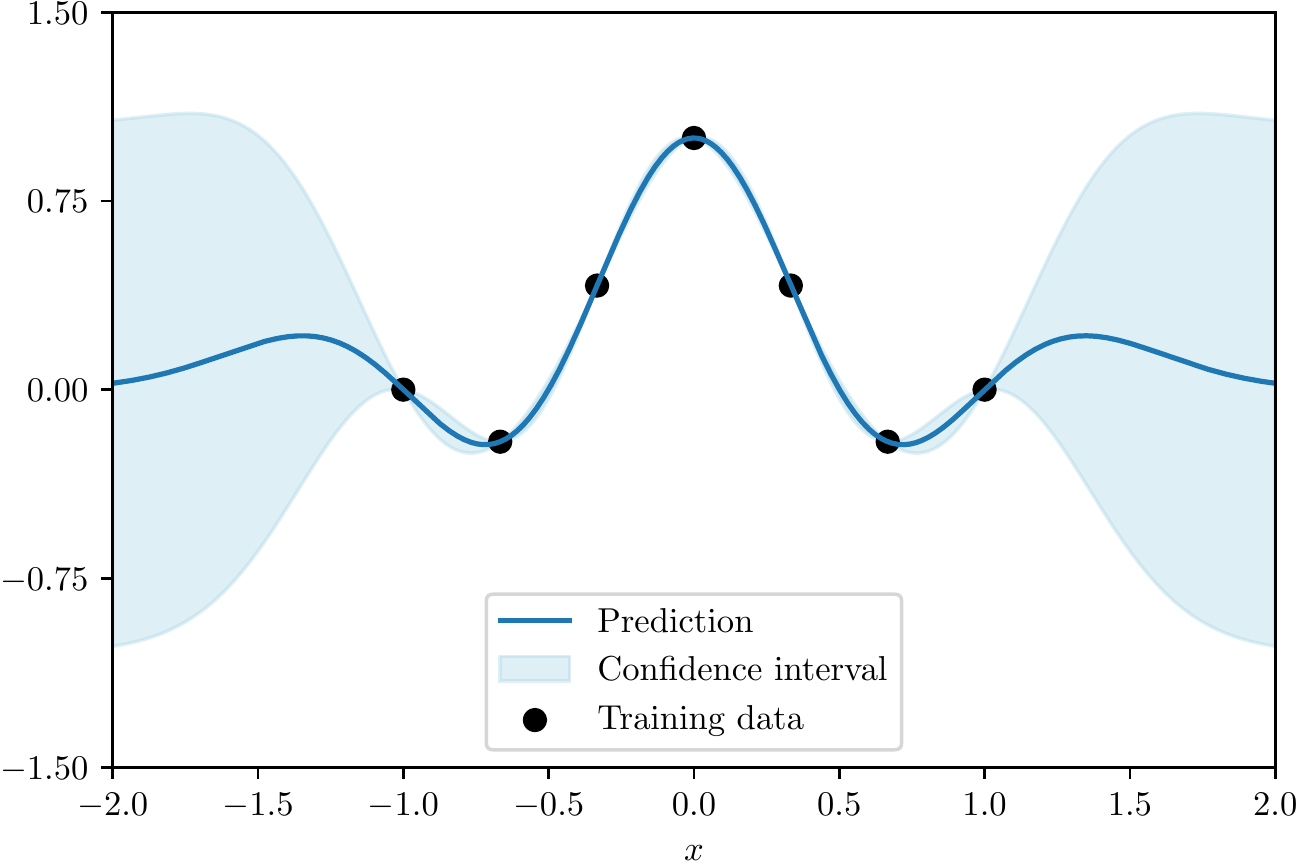}
\end{figure}

\begin{figure}[tbph]
\centering
\caption{Posterior distribution of $\func{sinc}\left(2x\right)$ ($\mathcal{K}_{\mathrm{Matern52}}$ kernel)}
\label{fig:gpr2b}
\figureskip
\includegraphics[width = \figurewidth, height = \figureheight]{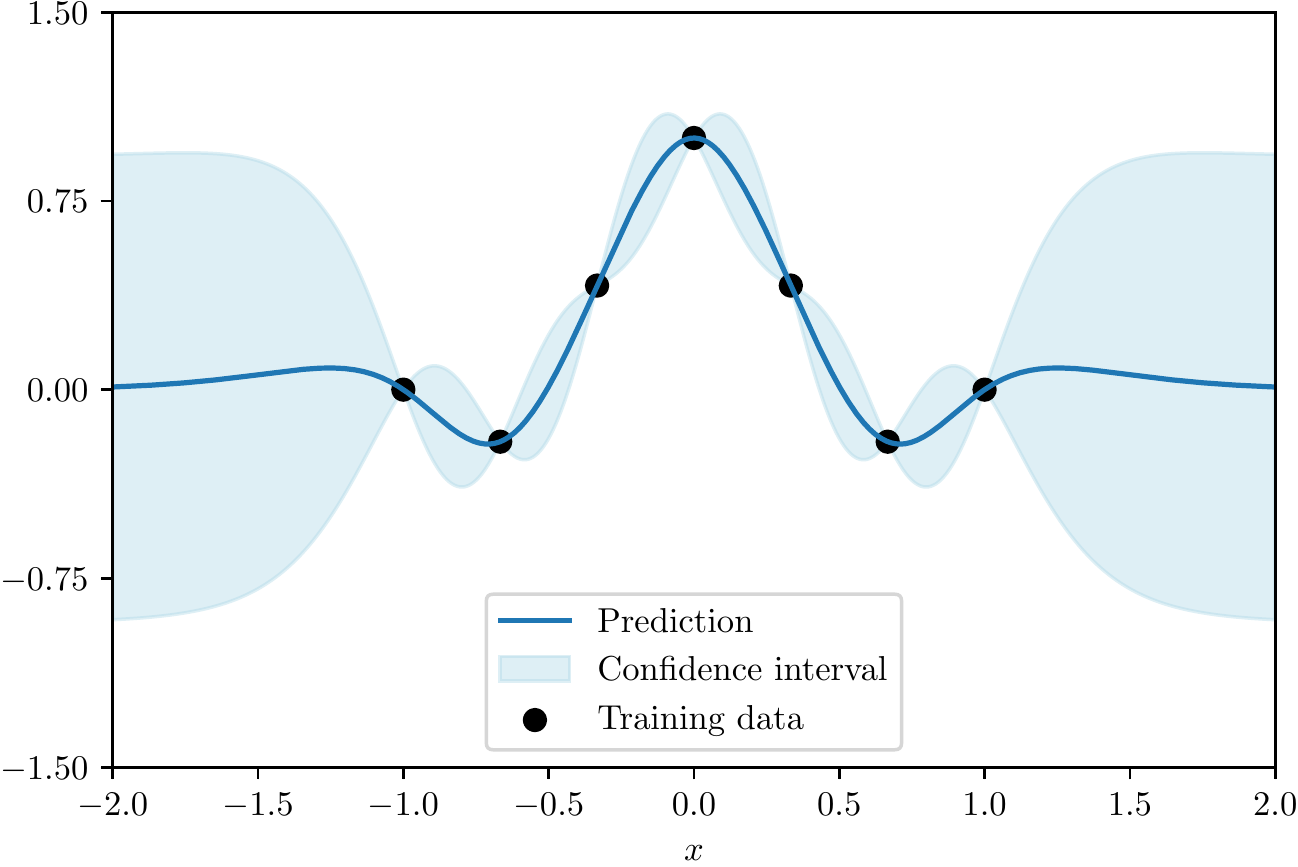}
\end{figure}

In Figures \ref{fig:gpr2a} and \ref{fig:gpr2b}, we consider a training set of
$7$ patterns measured without noise from the cardinal sine function
$\func{sinc}\left(2x\right)$ and we report the corresponding posterior
distribution $\hat{f}\left(x^{\star}\right)$ for two different kernels. The
blue solid line corresponds to the prediction, whereas the shaded area shows
the $95\%$
confidence interval of the forecast%
\footnote{It is equal to $\pm 2$ times the posterior standard deviation.}. We
notice that the square exponential kernel produces a lower interpolation
variance (Figure \ref{fig:gpr2a}) than the Matern32 kernel (Figure
\ref{fig:gpr2b}), whereas the extrapolation variance is similar for the two
kernel functions.

\paragraph{Periodic exponential kernel}

When working with time series, it is useful to be able to include periodicity
effects. Some covariance kernels exhibit this property, like the periodic
exponential kernel (MacKay, 1998):
\begin{equation*}
\mathcal{K}_{\mathrm{PE}}\left( x,x^{\prime }\right) =\sigma ^{2}\exp \left(
-\frac{1}{2}\sum_{j=1}^{d}\frac{1}{\ell _{j}^{2}}\sin ^{2}\left( \frac{\pi }{%
\lambda _{j}}\left( x_{j}-x_{j}^{\prime }\right) \right) \right)
\end{equation*}%
where each dimension of the input space has a period $\lambda _{j}$. More
periodic kernels can be computed from other kernels, such as the Mat\'ern class.
They can actually be built from any kernel for which the Gram matrix on a
periodic basis can be computed (Durrande, 2016).

\begin{remark}
Since kernels can be combined in various ways, we will often be interested in
separating time patterns from space patterns. If one observation is
defined by the couple $\left( t,x\right) $ where $x\in \mathbb{R}^{d}$ and $%
t\in \mathbb{R}$ is the time index\footnote{%
For example, we may consider daily prices of several assets.}, we can then
define a kernel on the whole space-time (Osborne \textsl{et al.}, 2012):
\begin{equation*}
\mathcal{K}\left( \left( x,t\right) ,\left( x^{\prime },t^{\prime }\right)
\right) =\mathcal{K}_{\textrm{Time}}\left( t,t^{\prime }\right) \cdot \mathcal{%
K}_{\textrm{Space}}\left( x,x^{\prime }\right)
\end{equation*}%
where $\mathcal{K}_{\textrm{Time}}$ is the kernel function for time patterns
and $\mathcal{K}_{\textrm{Space}}$ is the kernel function for space patterns.
\end{remark}

\paragraph{Spectral mixture kernel}

Wilson and Adams (2013) introduce a new kernel construction method based on
Gaussian mixtures in the Fourier space. For that, they use the Bochner's
theorem, which states that a real-valued function $k$ defined on
$\mathbb{R}^{d}$ is a covariance kernel of a stationary continuous random
process if and only if it can be represented in the following way:
\begin{equation*}
k\left( s\right) =\int_{\mathbb{R}^{d}}e^{2\pi i\lambda ^{\top }s}\,\mu
\left( \mathrm{d}\lambda \right)
\end{equation*}%
where $\mu $ is a positive finite measure on $\mathbb{R}^{d}$. This theorem
establishes equivalence between stationary covariance kernels\footnote{They
verify $\mathcal{K}\left( x,x^{\prime }\right) =k\left( x-x^{\prime }\right)
$.} and their Fourier transform. It is a generalization of the classic
one-dimensional spectral analysis when dealing with kernel functions instead of
autocovariance functions\footnote{This is why it is essential that the
covariance kernel is stationary.}. Indeed, the spectral density function
$f_{k}\left( \lambda \right) $ is the Fourier transform of the covariance
kernel function:
\begin{equation*}
f_{k}\left( \lambda \right) =\int_{\mathbb{R}^{d}}k\left( s\right) e^{-2\pi
i\lambda ^{\top }s}\,\mathrm{d}s
\end{equation*}%
whereas the covariance kernel function is the inverse Fourier transform of the
spectral density function $f_{k}\left( \lambda \right) $:
\begin{equation*}
k\left( s\right) =\int_{\mathbb{R}^{d}}f_{k}\left( \lambda \right) e^{2\pi
i\lambda ^{\top }s}\,\mathrm{d}\lambda
\end{equation*}%
For instance, the SE kernel has a spectral density which is Gaussian. This leads
Wilson and Adams (2013) to consider Gaussian mixtures of spectral densities to
extend the SE kernel.\smallskip

Let us define a mixture of $n_{m}$ Gaussian densities on $\mathbb{R}^{d}$ with
mean vectors $\left( \mu _{1},\ldots ,\mu _{n_{m}}\right) $ and diagonal
covariance matrices $\left( \Sigma _{1},\ldots ,\Sigma _{n_{m}}\right) $. The
corresponding density function $g\left( x\right) $ is defined by:
\begin{equation}
g\left( x\right) =\sum_{m=1}^{n_{m}}\omega _{m}\frac{1}{\left( {2\pi }%
\right) ^{d/2}\sqrt{\func{det}\Sigma _{m}}}\exp \left( -\left( x-\mu
_{m}\right) ^{\top }\Sigma _{m}^{-1}\left( x-\mu _{m}\right) \right)
\label{eq:mixture1}
\end{equation}%
where $\omega _{m}$ is the weight of the $m^{\mathrm{th}}$ Gaussian
distribution. In our case, we are interested in real-valued covariance
functions, implying that we replace $g\left( x\right) $ by $\frac{1}{2}\left(
g\left( x\right) +g\left( -x\right) \right) $. Interestingly, the inverse
Fourier transform of Equation (\ref{eq:mixture1}) is analytically tractable and
is given by\footnote{The correct formula is given in Wilson (2015).}:
\begin{equation*}
k_{\mathrm{SM}}\left( s\right) =\sum_{m=1}^{n_{m}}\omega _{m}\cos \left(
2\pi s^{\top }\mu _{m}\right) \exp \left( -2\pi ^{2}s^{\top }\Sigma
_{m}s\right)
\end{equation*}%
Wilson and Adams (2013) show that the spectral mixture (SM) kernel can recover
the usual kernels (squared exponential, Mat\'ern, rational quadratic). Another
interesting property is that it can learn negative covariances, which is
essential when considering mean-reverting processes and contrarian trading
strategies.

\subsubsection{Hyperparameter selection}

The covariance functions introduced before all have hyperparameters, such as
length scales $\Sigma =\limfunc{diag}\left( \ell _{1}^{2},\ldots ,\ell
_{d}^{2}\right) $\ in the squared exponential kernel, power $\alpha $ in the
rational quadratic kernel, etc. All these parameters influence how the GP model
can fit the observed data. This is why their choice is critical. They can be
fixed ex-ante or we can estimate them.\smallskip

For a given model, we denote by $\theta $ the parameters of the model. The
usual way of selecting parameters is to maximize the likelihood function
$L\left( \theta \right) =p\left( y\mid \theta \right) $. The underlying idea is
to maximize the probability of the sample data $y$. In the case of the Gaussian
process regression, $\theta =\left( \theta _{\mathcal{K}},\sigma _{\varepsilon
}\right) $ consists of the parameters $\theta _{\mathcal{K}}$ of the kernel
function and the standard deviation $\sigma _{\varepsilon }$ of the
noise. Let $z=f\left( x\right) $ be the GP. We have:%
\begin{equation*}
p\left( y\mid \theta \right) =\int p\left( y\mid \theta ,z\right) p\left(
z\mid \theta \right) \,\mathrm{d}z
\end{equation*}%
It is common to maximize the log marginal likelihood where we integrate out the
latent values of $z$ and solve:
\begin{equation*}
\hat{\theta}=\arg \max_{\theta }\LogL\left( \theta \right)
\end{equation*}%
where $\LogL\left( \theta \right) =\ln p\left( y\mid \theta \right) $. In
the case of Gaussian noise, we have $Y\sim \mathcal{N}\left( \mathbf{0}_{n},%
\mathcal{K}\left( \theta _{\mathcal{K}}\right) +\sigma _{\varepsilon
}^{2}I_{n}\right) $ where $\mathcal{K}\left( \theta _{\mathcal{K}}\right) $
denotes the kernel matrix that depends on the kernel parameters $\theta _{%
\mathcal{K}}$. It follows that:%
\begin{equation*}
\LogL\left( \theta \right) =-\frac{n}{2}\ln \left( 2\pi \right) -\frac{1}{2}%
\ln \left\vert \mathcal{K}\left( \theta _{\mathcal{K}}\right) +\sigma
_{\varepsilon }^{2}I_{n}\right\vert -\frac{1}{2}Y^{\top }\left( \mathcal{K}%
\left( \theta _{\mathcal{K}}\right) +\sigma _{\varepsilon }^{2}I_{n}\right)
^{-1}Y
\end{equation*}%
This problem is usually solved using gradient-descent or quasi-Newton
algorithms since it is possible to compute analytically the gradient of $%
\mathcal{K}\left( \theta _{\mathcal{K}}\right) $. However, $\LogL\left( \theta
\right) $ is not always convex, and may suffer local maxima (Duvenaud,
2013).\smallskip

\begin{figure}[tbph]
\centering
\caption{Posterior distribution before marginal likelihood maximization}
\label{fig:hyperparameter1}
\figureskip
\includegraphics[width = \figurewidth, height = \figureheight]{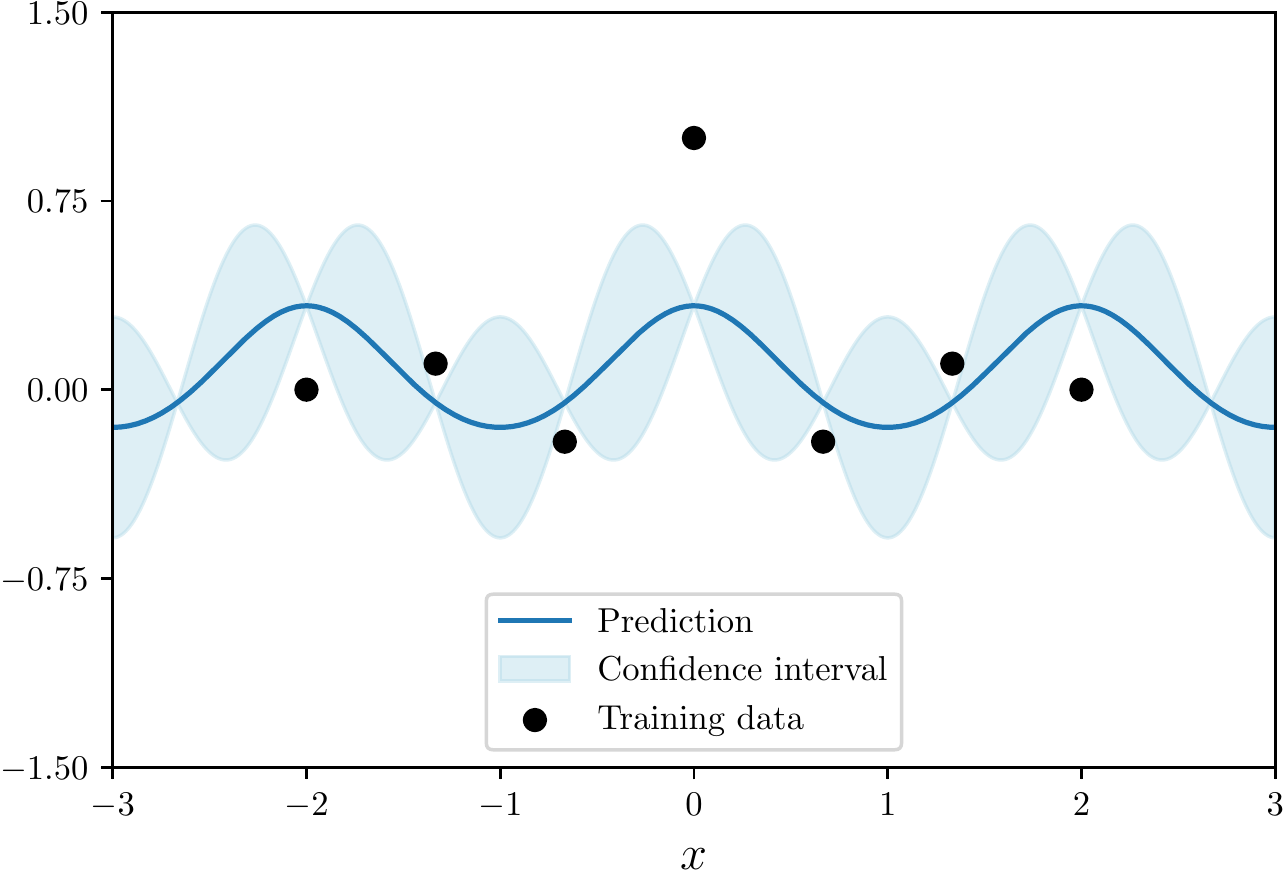}
\end{figure}

\begin{figure}[tbph]
\centering
\caption{Posterior distribution after marginal likelihood maximization}
\label{fig:hyperparameter2}
\figureskip
\includegraphics[width = \figurewidth, height = \figureheight]{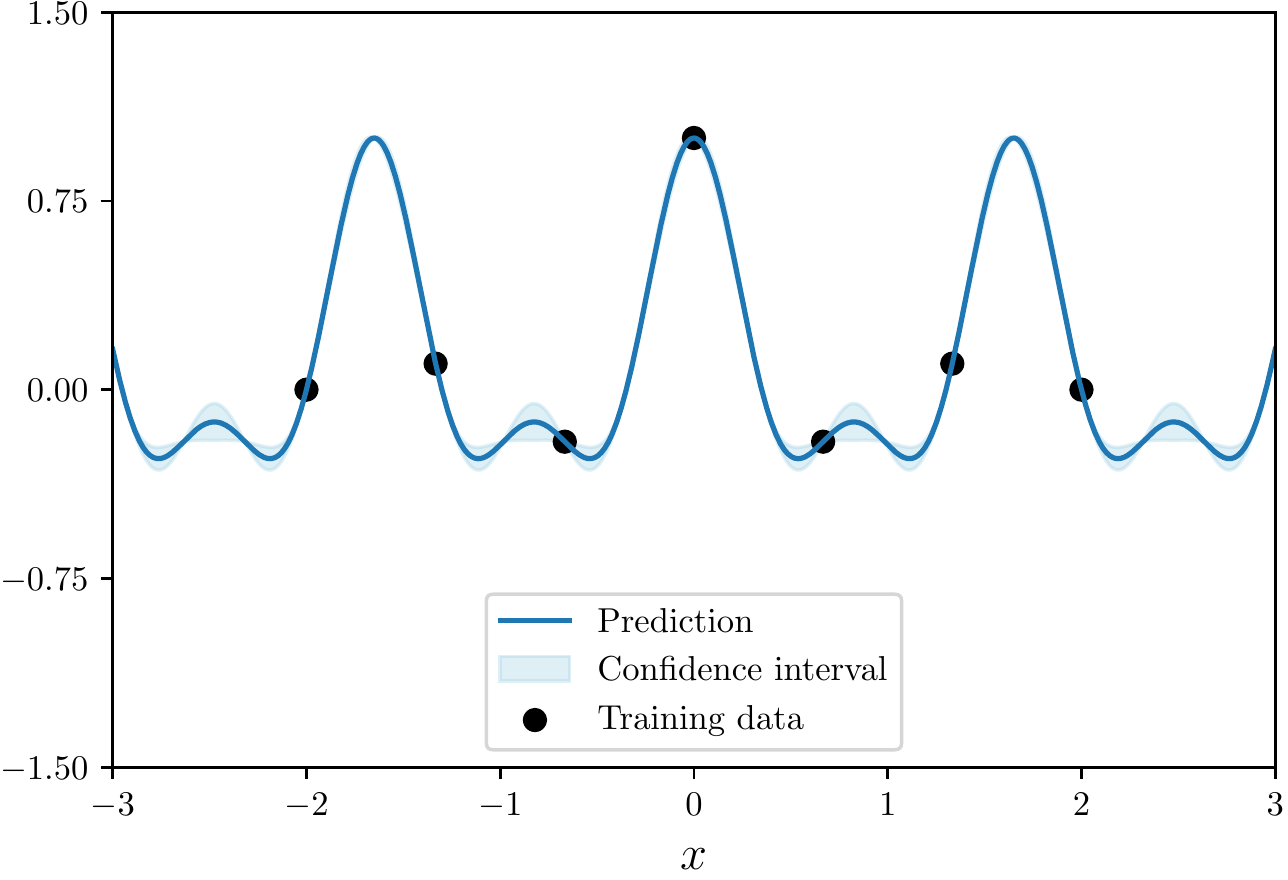}
\end{figure}

Let us illustrate the ML estimation of the hyperparameters with the periodic
kernel. For that, we use $7$ training points. In Figure
\ref{fig:hyperparameter1}, we report the posterior distribution $f\left(
x^{\star }\mid x,y\right) $ when $x^{\star }$ ranges from $-3$ to $+3$. We
assume that the hyperparameters of the kernel function are $\sigma =1$,
$\lambda _{1}=2$ and $\ell _{1}=1$ whereas the standard deviation of the noise
$\sigma _{\varepsilon }$ is set to $10^{-7}$. Then, we estimate the parameters
$\theta =\left( \sigma ,\lambda _{1},\ell _{1},\sigma _{\varepsilon }\right) $
by the method of maximum likelihood. We obtain $\hat{\sigma}=0.7657$,
$\hat{\lambda}_{1}=1.6506$, $\hat{\ell}_{1}=0.7664$ and $\sigma _{\varepsilon
}=2.46\times 10^{-7}$. The corresponding posterior distribution is given in
Figure \ref{fig:hyperparameter2}. As expected, we better fit the training set
after the ML estimation than before.

\begin{remark}
The Bayesian approach puts a prior distribution on the hyperparameters $\theta
$ and marginalizes the posterior GP distribution over $\theta$. However, this
is not analytically tractable. We also notice that the posterior distribution
of the GP is intractable if the noise is non-Gaussian. Both cases
require to use Monte Carlo methods such as the Hamiltonian or Hybrid Monte
Carlo (Neal, 2011), which is described in Appendix
\ref{appendix:hybrid-monte-carlo} on page
\pageref{appendix:hybrid-monte-carlo}.
\end{remark}

\subsubsection{Classification}

Gaussian processes regression can be extended for classification problems,
where the output is a discrete variable corresponding to the class index. For
example, we could want to predict the movements of asset prices: $1$ for a
positive return and $0$ otherwise. In what follows, we consider the case of
binary classification.\smallskip

To model two classes with a GP prior over the data, one generally uses a
sigmoid function\footnote{This means that $g\left( x\right) $ is a
monotonically increasing function in $\left[ 0,1\right] $.} $g\left( x\right)
$, for example the logistic function $\func{logit}\left( x\right) =\left(
1+e^{-x}\right) ^{-1}$. The output $y$ is such that:
\begin{equation*}
\Pr \left\{ y=1\right\} =g\left( f\left( x\right) \right)
\end{equation*}%
where $f\left( x\right) $ is the GP over $\mathcal{X}$. The predictive
distribution for new inputs $x^{\star }$ can be marginalized over the latent
GP values:%
\begin{equation*}
p\left( z^{\star }\mid y,z\right) =\int p\left( z^{\star }\mid z\right)
p\left( z\mid y\right) \,\mathrm{d}z
\end{equation*}
where $z$ and $z\mathbf{^{\star }}$ respectively denote the random variables
$f\left( x\right) $ and $f\left( x^{\star }\right) $. We deduce that:
\begin{equation}
\Pr \left\{ y^{\star }=1\mid y\right\} =\int g\left( z^{\star }\right)
p\left( z^{\star }\mid z\right) p\left( z\mid y\right) \,\mathrm{d}z^{\star
}\,\mathrm{d}z  \label{eq:gpc1}
\end{equation}%
where the posterior distribution $p\left( z\mid y\right) $ can be written using
Bayes rule:
\begin{equation*}
p\left( z\mid y\right) =\frac{p\left( y\mid z\right) p\left( z\right) }{%
p\left( y\right) }
\end{equation*}%
Here, $p\left( z^{\star }\mid z\right) $ is the usual posterior distribution of
the GP. However, the posterior distribution $p\left( z\mid y\right) $ is not
easy to compute, and this is why approximations are used to evaluate the
integral (\ref{eq:gpc1}). Laplace approximation and expectation propagation are
the two popular methods (Rasmussen and Williams, 2006). The first one
approximates the posterior using a two-order Taylor expansion around its
maximum, while the second one approximates the intractable probability
distribution by minimizing the Kullback-Leibler divergence.

\subsection{Bayesian optimization}

Bayesian optimization is a black-box optimization method, meaning that little
information is known about the objective function $f\left(x\right)$. Typically,
Bayesian optimization is useful when the function is expensive to evaluate, its
analytical expression is inaccessible or the gradient vector is not stable.
This is the case with many complex machine learning problems where one would
like to optimize the hyperparameters. For instance, the score of a deep neural
network architecture is difficult to compute for a given set of hyperparameters
(because training the model itself can take a long time), and it is impossible
to compute the gradient vector with respect to each hyperparameter.

\subsubsection{General principles}

We are interested in finding the maximum of $f\left( x\right) $ on some bounded
set $\mathcal{X}$. Bayesian optimization consists of two parts: (1) the
\textquoteleft \textit{probabilistic surrogate}\textquoteright\ and (2) the
\textquoteleft \textit{acquisition function}\textquoteright\ (or utility
function). First of all, we build a prior probabilistic model for the objective
function $f\left( x\right) $, and then update the probability distribution with
samples drawn from $f\left( x\right) $ to get a posterior probability
distribution. This approximation of the objective function is called a
surrogate model. Gaussian processes are a popular surrogate model for Bayesian
optimization because the GP posterior is still a multivariate normal
distribution\footnote{Other models exist such as random forests (Hutter
\textsl{et al.}, 2011).}. We then use a utility function based on this
posterior probability distribution to choose a new point to evaluate the
objective function at the next step. This utility function is called
acquisition function. Intuitively, we consider the trade-off between
exploitation and exploration. Exploitation means sampling where the surrogate
model predicts a high objective gain and exploration means sampling where the
prediction uncertainty is high. Therefore, the general idea of Bayesian
optimization consists of the following steps:
\begin{enumerate}
\item Place a GP prior on the objective function $f\left( x\right) $.

\item Update the GP posterior probability distribution on $f\left( x\right) $
    with all available samples.

\item Based on the acquisition function, decide where to make the next
    measurement.

\item Given this measurement, update the GP posterior probability
    distribution.

\item Repeat steps 2-4 until an approximated maximum of the objective
    function $f\left( x\right) $ is obtained (or stop after a predefined
    number of iterations).
\end{enumerate}

\subsubsection{Acquisition function}

We assume that the function $f\left( x\right) $ has a Gaussian process prior
and we observe samples of the form $\left\{ \left( x_{i},y_{i}\right) \right\}
_{i=1}^{n}$. We have $y_{i}=f\left( x_{i}\right) +\varepsilon _{i}$ where
$\varepsilon _{i}\sim \mathcal{N}\left( 0,\sigma _{\varepsilon }^{2}\right) $
is the noise process. We denote by $x $ and $y$ the matrices $\left(
x_{1},\dots ,x_{n}\right) $ and $\left( y_{1},\dots ,y_{n}\right) $. As shown
previously, we can compute the posterior probability distribution $f\left(
x^{\star }\mid x,y\right) $ for a new observation $x^{\star }$, and we have:
\begin{equation*}
\hat{f}_{n}\left( x^{\star }\right) \sim \mathcal{N}\left( \hat{m}_{n}\left(
x^{\star }\right) ,\mathcal{\hat{K}}_{n}\left( x^{\star },x^{\star }\right)
\right)
\end{equation*}%
where:%
\begin{equation*}
\hat{m}_{n}\left( x^{\star }\right) =\mathcal{K}\left( x^{\star },x\right)
\left( \mathcal{K}\left( x,x\right) +\sigma _{\varepsilon }^{2}I_{n}\right)
^{-1}y
\end{equation*}%
and:%
\begin{equation*}
\mathcal{\hat{K}}_{n}\left( x^{\star },x^{\star }\right) =\mathcal{K}\left(
x^{\star },x^{\star }\right) -\mathcal{K}\left( x^{\star },x\right) \left(
\mathcal{K}\left( x,x\right) +\sigma _{\varepsilon }^{2}I_{n}\right) ^{-1}%
\mathcal{K}\left( x,x^{\star }\right)
\end{equation*}%
The subscript $n$ indicates that $\hat{f}_{n}$, $\hat{m}_{n}$ and $\mathcal{%
\hat{K}}_{n}$ depend on the sample of size $n$, which corresponds to the
optimization step $n$. We note $\mathcal{D}_{n}$ the augmented data with the
GP:
\begin{equation*}
\mathcal{D}_{n}=\left\{ \left( x_{i},y_{i},\hat{f}_{i}\left( x_{i}\right)
\right) \right\} _{i=1}^{n}
\end{equation*}%
Let $\mathcal{U}_{n}\left( x^{\star }\right) $ be the acquisition (or utility)
function based on $\mathcal{D}_{n}$. The Bayesian optimization
consists then in finding the new optimal point $x_{n+1} \in \mathcal{X}$ such that:%
\begin{equation*}
x_{n+1}=\arg \max\, \mathcal{U}_{n}\left( x^{\star }\right)
\end{equation*}%
and updating the set of observations and the posterior distribution (see
Algorithm \ref{alg:algorithm1}).

\begin{algorithm}[tbph]
\caption{Bayesian optimization algorithm}
\label{alg:algorithm1}
\begin{algorithmic}
\STATE The goal is to perform a Bayesian optimization
\STATE We initialize the data sample $\mathcal{D}_{1}$ and the vector $\theta_1$ of hyperparameters
\FOR {$n = 1, 2, \ldots$}
    \STATE Find the optimal value $x_{n+1}\in \mathcal{X}$ of the utility maximization problem:
        \begin{equation*}
            x_{n+1}=\arg \max\, \mathcal{U}_{n}\left( x^{\star }\right)
        \end{equation*}
    \STATE Update the data:%
        \begin{equation*}
            \mathcal{D}_{n+1}\leftarrow \mathcal{D}_{n}\cup \left\{ \left(x_{n+1},y_{n+1},\hat{f}_{n+1}\left( x_{n+1}\right) \right) \right\}
        \end{equation*}
    \STATE Update the hyperparameter vector $\theta_{n+1}$ of the kernel function
\ENDFOR
\RETURN $\mathcal{D}_{n}$ and $\theta_{n}$
\end{algorithmic}
\end{algorithm}

\paragraph{Improvement-based acquisition function}

Let $f_{n}\left( \varkappa _{n}^{\star }\right) $ be the current optimal
value among $n$ samples drawn from $f\left( x\right) $:%
\begin{equation*}
\varkappa _{n}^{\star }=\arg \max_{\varkappa \in x}f\left( \varkappa \right)
\end{equation*}%
where $\varkappa _{n}^{\star }$ is the point that maximizes the GP function
over the first $n$ steps. We would like to choose the next point $x_{n+1}$
to be evaluated in order to improve this value. We define the improvement $%
\Delta _{n}\left( x^{\star }\right) $ as follows:
\begin{eqnarray*}
\Delta _{n}\left( x^{\star }\right)  &=&\left( \hat{f}_{n}\left( x^{\star
}\right) -f_{n}\left( \varkappa _{n}^{\star }\right) \right) ^{+} \\
&=&\max \left( \hat{f}_{n}\left( x^{\star }\right) -f_{n}\left( \varkappa
_{n}^{\star }\right) ,0\right)
\end{eqnarray*}%
The most intuitive strategy, as proposed by Kushner (1964), is to choose the
point that maximizes the probability of a positive improvement:
\begin{eqnarray*}
\Pr \left\{ \Delta _{n}\left( x^{\star }\right) >0\right\}  &=&\Pr \left\{
\hat{f}_{n}\left( x^{\star }\right) >f_{n}\left( \varkappa _{n}^{\star
}\right) \right\}  \\
&=&\Pr \left\{ \mathcal{N}\left( 0,1\right) >\frac{f_{n}\left( \varkappa
_{n}^{\star }\right) -\hat{m}_{n}\left( x^{\star }\right) }{\sqrt{\mathcal{%
\hat{K}}_{n}\left( x^{\star },x^{\star }\right) }}\right\}  \\
&=&\Phi \left( \frac{\hat{m}_{n}\left( x^{\star }\right) -f_{n}\left(
\varkappa _{n}^{\star }\right) }{\sqrt{\mathcal{\hat{K}}_{n}\left( x^{\star
},x^{\star }\right) }}\right)
\end{eqnarray*}%
Since the probability of improvement fails to quantify the level of
improvement, Mo\v{c}kus (1975) introduces an alternative acquisition function
which takes into account the expected value of improvement (EI):
\begin{equation*}
\limfunc{EI}\nolimits_{n}\left( x^{\star }\right) =\mathbb{E}\left[ \Delta
_{n}\left( x^{\star }\right) \right]
\end{equation*}%
In the GP framework, we obtain a closed form of the expected improvement
acquisition function\footnote{See Appendix \ref{appendix:black-formula} on page
\pageref{appendix:black-formula}.}:
\begin{eqnarray*}
\limfunc{EI}\nolimits_{n}\left( x^{\star }\right)  &=&\left( \hat{m}%
_{n}\left( x^{\star }\right) -f_{n}\left( \varkappa _{n}^{\star }\right)
\right) \Phi \left( \frac{\hat{m}_{n}\left( x^{\star }\right) -f_{n}\left(
\varkappa _{n}^{\star }\right) }{\sqrt{\mathcal{\hat{K}}_{n}\left( x^{\star
},x^{\star }\right) }}\right) + \\
&&\sqrt{\mathcal{\hat{K}}_{n}\left( x^{\star },x^{\star }\right) }\phi
\left( \frac{\hat{m}_{n}\left( x^{\star }\right) -f_{n}\left( \varkappa
_{n}^{\star }\right) }{\sqrt{\mathcal{\hat{K}}_{n}\left( x^{\star },x^{\star
}\right) }}\right)
\end{eqnarray*}%
It follows that $\func{EI}_{n}\left( x^{\star }\right) $ and its derivatives
are easy to evaluate, and we can use optimization algorithms such as
quasi-Newton methods to find its maximum. Here, we have defined two utility
functions $\mathcal{U}_{n}\left( x^{\star }\right) =\Pr \left\{ \Delta
_{n}\left( x^{\star }\right) >0\right\} $ and $\mathcal{U}_{n}\left( x^{\star
}\right) =\func{EI}_{n}\left( x^{\star }\right) $ that are good candidates of
acquisition functions. Applications of improvement-based acquisition functions
are studied in Jones \textsl{et al.} (1998), Jones (2001), Brochu \textsl{et
al.} (2010), and Shahriari \textsl{et al.} (2016), whereas the convergence of
improvement-based optimization has been shown by Bull (2011). Appendix
\ref{appendix:improvement-minimization} on page
\pageref{appendix:improvement-minimization} extends the previous results to
minimization problems.

\begin{remark}
The previous approach can be generalized by considering a given threshold $\tau
$. In this case, we define the improvement by $\Delta _{n}\left( x^{\star
}\right) =\left( \hat{f}_{n}\left( x^{\star }\right) -\tau \right)
^{+}$. We have:%
\begin{equation*}
\Pr \left\{ \Delta _{n}\left( x^{\star }\right) >0\right\} =\Phi \left(
\frac{\hat{m}_{n}\left( x^{\star }\right) -\tau }{\sqrt{\mathcal{\hat{K}}%
_{n}\left( x^{\star },x^{\star }\right) }}\right)
\end{equation*}%
and:%
\begin{equation*}
\limfunc{EI}\nolimits_{n}\left( x^{\star }\right) =\left( \hat{m}_{n}\left( x^{\star
}\right) -\tau \right) \Phi \left( \frac{\hat{m}_{n}\left( x^{\star }\right)
-\tau }{\sqrt{\mathcal{\hat{K}}_{n}\left( x^{\star },x^{\star }\right) }}%
\right) +\sqrt{\mathcal{\hat{K}}_{n}\left( x^{\star },x^{\star }\right) }%
\phi \left( \frac{\hat{m}_{n}\left( x^{\star }\right) -\tau }{\sqrt{\mathcal{%
\hat{K}}_{n}\left( x^{\star },x^{\star }\right) }}\right)
\end{equation*}%
Most of the time, the threshold $\tau $ is set to $f_{n}\left( \varkappa
_{n}^{\star }\right) +\xi $ where $\xi >0$.
\end{remark}

\begin{figure}[tbph]
\centering
\caption{Objective function of the minimization problem}
\label{fig:acquisition1}
\figureskip
\includegraphics[width = \figurewidth, height = \figureheight]{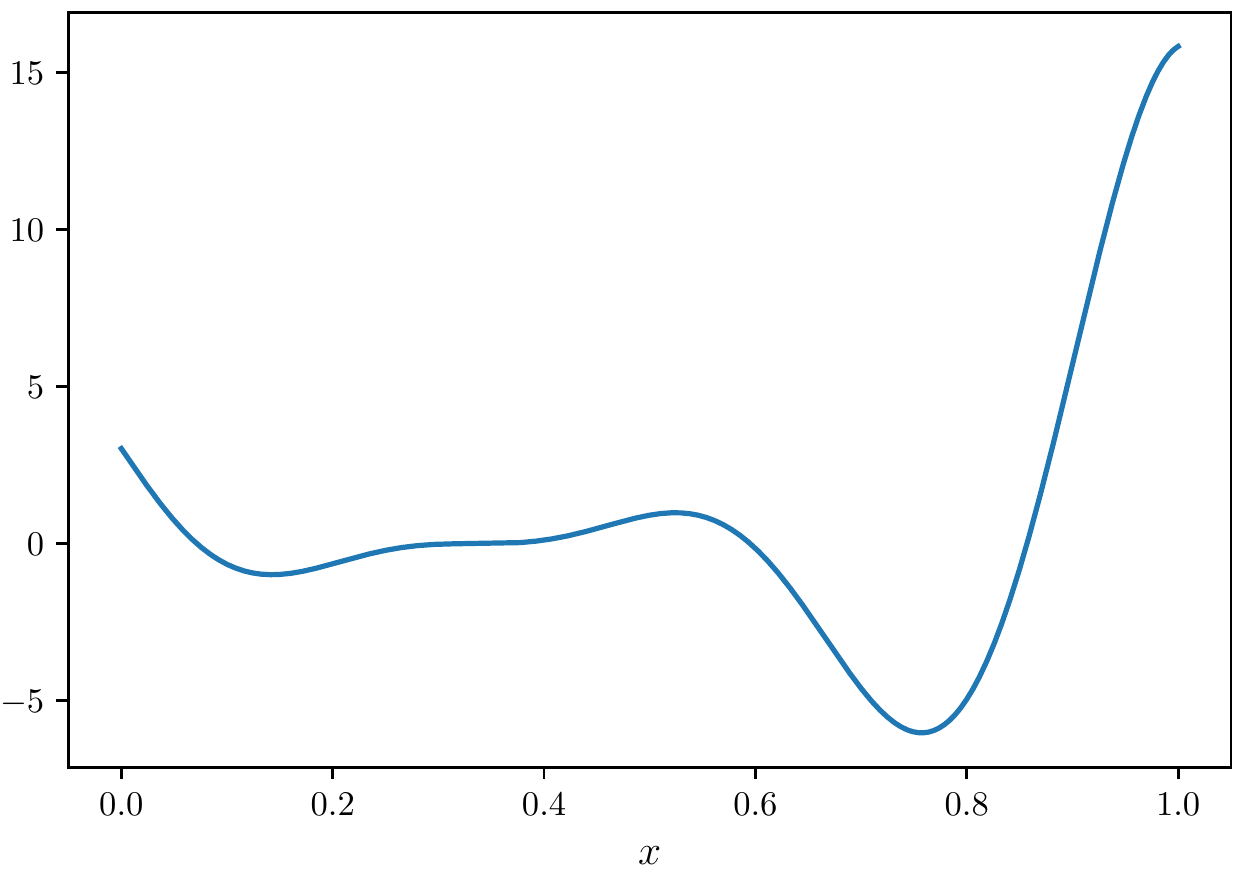}
\end{figure}

\begin{figure}[tbph]
\centering
\caption{Iterations of the Bayesian optimization}
\label{fig:acquisition2}
\figureskip
\includegraphics[width=\linewidth,height=18cm,keepaspectratio]{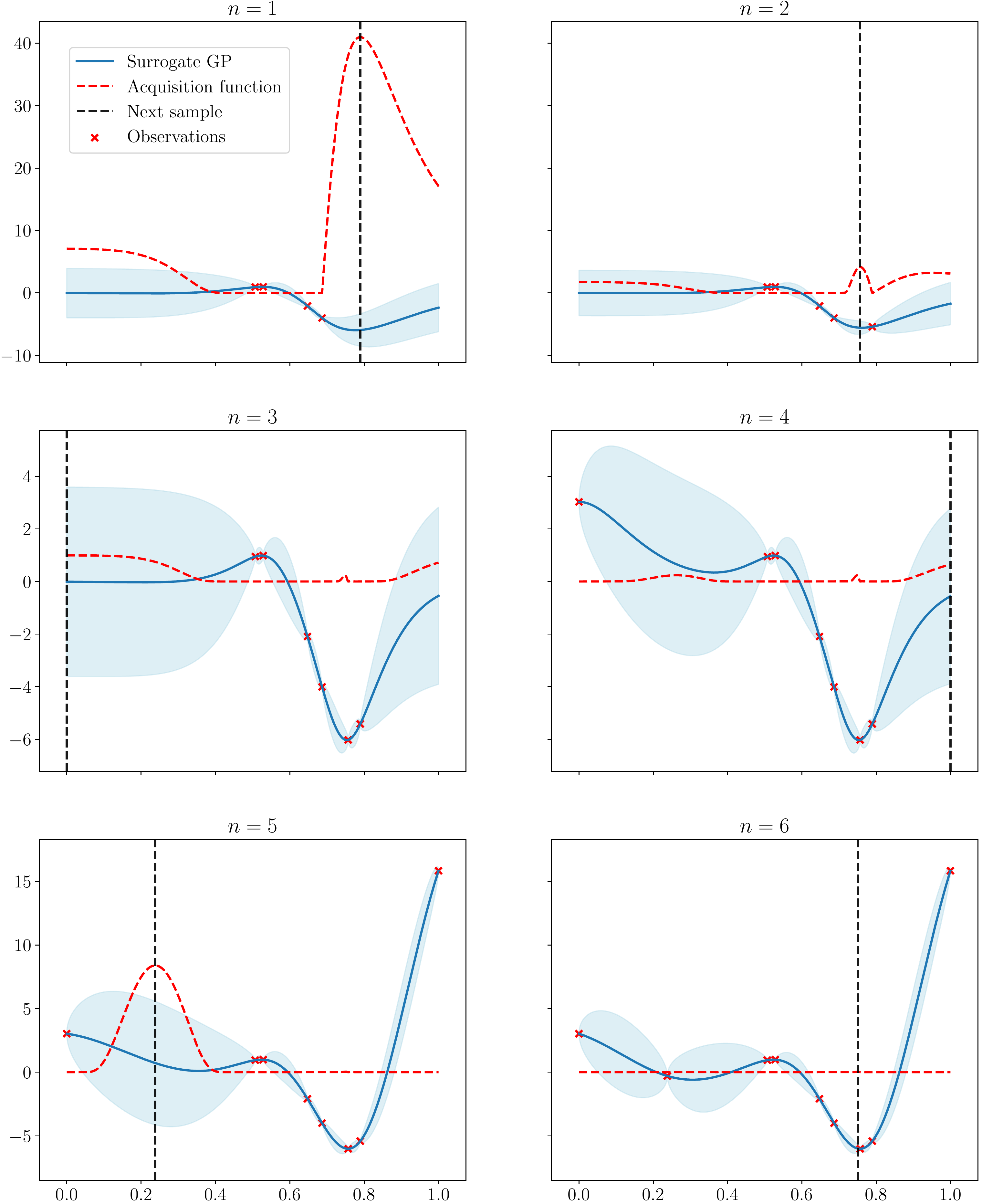}
\end{figure}

In Figure \ref{fig:acquisition2}, we illustrate the improvement-based
optimization using the following minimization problem\footnote{This example is
taken from Forrester \textsl{et al.} (2008).}:
\begin{equation*}
\min f\left( x\right) =\left( 6x-2\right) ^{2}\sin \left( 12x-4\right)
\end{equation*}%
The objective function $f\left( x\right) $ is reported in Figure
\ref{fig:acquisition1}. In practice, we start with an initial design, usually
consisting in measuring several random points of the domain. In the top/left
panel in Figure \ref{fig:acquisition2}, we start the algorithm with three
initial points. We report the mean (blue solid line) and the confidence
interval (blue shaded area) of the GP distribution. We also show the
acquisition function $\mathcal{U}_{n}\left( x^{\star }\right)
=\limfunc{EI}\nolimits_{n}\left( x^{\star }\right) $ (red dashed line) and
indicate the suggested next location by a vertical black line, which
corresponds to the maximum of $\mathcal{U}_{n}\left( x^{\star }\right) $. The
top/right panel corresponds to the second iteration where we have updated the
sample. Indeed, the sample now contains the initial three points and the
maximum point $x^{\star }$ obtained at the previous iteration. Then, we
continue the process and show the results of the Bayesian optimization for the
next five steps. We notice that steps $n=3$, $n=4$  and $n=5$ correspond to an
exploration stage (sampling where the variance is high), whereas step $n=1$,
$n=2$ and $n=6$ corresponds to an exploitation stage (sampling where the
improvement is high). Finally, after six iterations, we have located the
minimum since the acquisition function is equal to zero.

\paragraph{Entropy-based acquisition function}

In this paragraph, we introduce two other acquisition functions based on
differential entropy of information theory. Given random variables $X$ and $Y$
with continuous probability functions $p\left( x\right) $ and $p\left( y\right)
$ and joint probability function $p\left( x,y\right) $, the differential (or
Shannon) entropy $H\left( X\right) $ is equal to:
\begin{equation*}
H\left( X\right) =-\int p\left( x\right) \ln p\left( x\right) \,\mathrm{d}x
\end{equation*}%
while the conditional differential entropy $H\left( X\mid Y\right) $ of $X$ is
defined by:
\begin{equation*}
H\left( X\mid Y\right) =-\int \int p\left( x,y\right) \ln p\left( x\mid
y\right) \,\mathrm{d}x\,\mathrm{d}y
\end{equation*}%
Using Bayes theorem, we note that $H\left( X\mid Y\right) $ is the result of
averaging $H\left( X\mid Y=y\right) $ over all possible values $y$ of the
random variable $Y$ (Cover and Thomas, 2012).\smallskip

We consider the location $x_{\max }$ of the global maximum of $f\left( x\right)
$ as a random variable, which has the posterior distribution $\hat{p}_{n}\left(
x_{\max }\right) =p\left( x_{\max }\mid x,y\right) $. Then, we can use the
differential entropy to quantify the uncertainty of this point. The smaller the
differential entropy, the lower the uncertainty. In order to have more certainty
about the location of the global minimum, we want to choose the next point to
evaluate $x_{n+1}$ that implies the largest decrease in the differential
entropy. For this purpose, we define the entropy search (ES) acquisition
function as the difference between the current differential entropy of
$\hat{p}_{n}\left( x_{\max }\right) $ and the expected differential entropy of
posterior probability distribution $p\left( x_{\max }\mid x,y,x^{\star
},\hat{f}_{n}\left( x^{\star }\right) \right) $ after adding a new sample
$\left\{ x^{\star },\hat{f}_{n}\left( x^{\star }\right) \right\} $:
\begin{equation*}
\limfunc{ES}\nolimits_{n}\left( x^{\star }\right) =H\left( x_{\max }\right)
-H\left( x_{\max }\mid \hat{f}_{n}\left( x^{\star }\right) \right)
\end{equation*}%
It follows that the next point is the solution of the maximization problem:
\begin{equation*}
x_{n+1}=\arg \max_{x^{\star }\in \mathcal{X}}\limfunc{ES}\nolimits_{n}\left(
x^{\star }\right)
\end{equation*}%
Although Henning and Schuler (2012) propose a method to approximate the above
equation, Frazier (2018) indicates some difficulties in practice:
\begin{itemize}
\item $\hat{p}_{n}\left( x_{\max }\right) $ does not have always a closed-form
    expression;
\item we need to compute the differential entropy of a large number of
    samples of $\left\{ x^{\star },\hat{f}_{n}\left( x^{\star }\right)
    \right\} $
to evaluate the expectation in the second term $H\left( x_{\max }\mid \hat{f}%
_{n}\left( x^{\star }\right) \right) $.
\end{itemize}
This is why Hern\'andez-Lobato \textsl{et al.} (2014) propose an alternative
approach called \textit{predictive entropy search} (PES):%
\begin{equation*}
\limfunc{PES}\nolimits_{n}\left( x^{\star }\right) =H\left( \hat{f}%
_{n}\left( x^{\star }\right) \right) -H\left( \hat{f}_{n}\left( x^{\star
}\right) \mid x_{\max }\right)
\end{equation*}
Using the symmetric property of the mutual information, we can demonstrate that
$\limfunc{PES}\nolimits_{n}\left( x^{\star }\right) $ and
$\limfunc{ES}\nolimits_{n}\left( x^{\star }\right) $ are equivalent acquisition
functions\footnote{We have:%
\begin{eqnarray*}
\limfunc{ES}\nolimits_{n}\left( x^{\star }\right)  &=&I\left( x_{\max },\hat{%
f}_{n}\left( x^{\star }\right) \right)  \\
&=&I\left( \hat{f}_{n}\left( x^{\star }\right) ,x_{\max }\right)  \\
&=&\limfunc{PES}\nolimits_{n}\left( x^{\star }\right)
\end{eqnarray*}%
where $I\left( X,Y\right) $ is the mutual information of two continuous
random variables:%
\begin{eqnarray*}
I\left( X,Y\right)  &=&\int \int p\left( x,y\right) \ln \frac{p\left(
x,y\right) }{p\left( x\right) p\left( y\right) }\,\mathrm{d}x\,\mathrm{d}y \\
&=&H\left( X\right) -H\left( X,Y\right)
\end{eqnarray*}%
}. In the case of the PES acquisition function, we can compute a closed-form
expression for $H\left( \hat{f}_{n}\left( x^{\star }\right) \right) $ and
Hern\'andez-Lobato \textsl{et al.} (2014) uses the expectation propagation
method (Minka, 2001) to find an approximation of $H\left( \hat{f}_{n}\left(
x^{\star }\right) \mid x_{\max }\right) $. Therefore, we can find the maximum
of the PES acquisition function by a simulation approach.

\paragraph{Knowledge gradient-based acquisition function}

The knowledge gradient (KG) acquisition function is closed to the expected
improvement. It was first introduced in Frazier \textsl{et al.} (2009) for
finite discrete decision spaces before Scott \textsl{et al.} (2011) extended it
to Gaussian processes. The main difference between KG and EI acquisition
functions is that KG accounts for noise and does not restrict the final
solution to a previously evaluated point, meaning that it can return any
point of the domain and not only one observed point.\smallskip

Suppose that we have observed the sample $\left\{ \left( x_{i},y_{i}\right)
\right\} _{i=1}^{n}$. As previously, we compute the posterior probability
distribution:
\begin{equation*}
\hat{f}_{n}\left( x^{\star }\right) \sim \mathcal{N}\left( \hat{m}_{n}\left(
x^{\star }\right) ,\mathcal{\hat{K}}_{n}\left( x^{\star },x^{\star }\right)
\right)
\end{equation*}
Under risk-neutrality assumption (Berger, 2013), we value a random outcome
according to its expected value $\hat{m}_{n}\left( x^{\star }\right) $. To
maximize the objective function, we can try to maximize $\hat{m}_{n}\left(
x^{\star }\right) $ and we note $\hat{m}_{n}\left( \varkappa _{n}^{\star
}\right) =\max_{\varkappa}\hat{m}_{n}\left( \varkappa \right) $. If we consider
one supplementary point, we could also compute the new posterior distribution
for $f$ with conditional expected value $\hat{m}_{n+1}\left( x^{\star }\right)
$. The idea is then to choose a new point that maximizes the increment in
conditioned expectation gained from sampling this point. For example, we can
maximize the expected value of the difference, which is called \textit{knowledge
gradient}:
\begin{equation*}
\limfunc{KG}\nolimits_{n}\left( x^{\star }\right) =\mathbb{E}\left[ \hat{m}%
_{n+1}\left( x^{\star }\right) -\hat{m}_{n}\left( \varkappa _{n}^{\star
}\right) \right]
\end{equation*}
We have:
\begin{equation*}
x_{n+1}=\arg \max_{x^{\star }\in \mathcal{X}}\limfunc{KG}\nolimits_{n}\left(
x^{\star }\right)
\end{equation*}%
The solution can then be obtained via simulation using Algorithm \ref{alg:algorithm2} formulated
by Frazier (2018).

\begin{algorithm}[tbph]
\caption{Simulation-based computation of $\limfunc{KG}\nolimits_{n}\left(x^{\star }\right)$}
\label{alg:algorithm2}
\begin{algorithmic}
\STATE $x^{\star }$ is the input parameter
\STATE $n_s$ is the number of simulations
\STATE We note $\hat{m}_{n}\left( \varkappa _{n}^{\star }\right) =\max_{\varkappa }\hat{m}_{n}\left( \varkappa \right) $
\FOR{$s=1:n_{s}$}
    \STATE Generate $y_{\left( s\right) }^{\star }\sim \mathcal{N}\left(
        \hat{m}_{n}\left( x^{\star }\right) ,\mathcal{\hat{K}}_{n}\left( x^{\star},x^{\star }\right) \right) $
    \STATE Add the simulated point $(x^{\star },y_{\left( s\right) }^{\star })$ to the current sample $\left( x,y\right) $
    \STATE Compute $\hat{m}_{n+1}\left( \varkappa _{\left( s\right) }\right)=\max_{\varkappa \in \mathcal{X}}\hat{m}_{n+1}\left( \varkappa \right) $
        where $\hat{m}_{n+1}\left( \varkappa \right) $ is the posterior mean that depends on $(x^{\star },y_{\left( s\right) }^{\star })$
    \STATE $\Delta _{\left( s\right) }\longleftarrow \hat{m}_{n+1}\left( \varkappa _{\left( s\right) }\right) -\hat{m}_{n}\left( \varkappa _{n}^{\star }\right)$
\ENDFOR
\RETURN $\limfunc{KG}\nolimits_{n}\left( x^{\star }\right) \leftarrow n_{s}^{-1}\sum_{s=1}^{n_{s}}\Delta _{\left( s\right) }$
\end{algorithmic}
\end{algorithm}

\begin{remark}
In the noise-free case and if the final solution is limited to the previous
sampling, the KG acquisition function reduces to the EI acquisition function:
\begin{eqnarray*}
\hat{m}_{n+1}\left( x^{\star }\right) -\hat{m}_{n}\left( \varkappa ^{\star
}\right)  &=&\max \left( f_{n}\left( \varkappa _{n}^{\star }\right) ,\hat{f}%
_{n}\left( x^{\star }\right) \right) -f_{n}\left( \varkappa _{n}^{\star
}\right)  \\
&=&\left( \hat{f}_{n}\left( x^{\star }\right) -f_{n}\left( \varkappa
_{n}^{\star }\right) \right) ^{+}
\end{eqnarray*}
\end{remark}

\section{Financial applications}

In this section, we use Gaussian processes to model and forecast the yield curve and Bayesian optimization
to build an online trend-following strategy.

\subsection{Yield curve modeling}

\begin{figure}[tbph]
\centering
\caption{GP fitting of the yield curve (June 2007)}
\label{fig:yieldcurve1}
\figureskip
\includegraphics[width = \figurewidth, height = \figureheight]{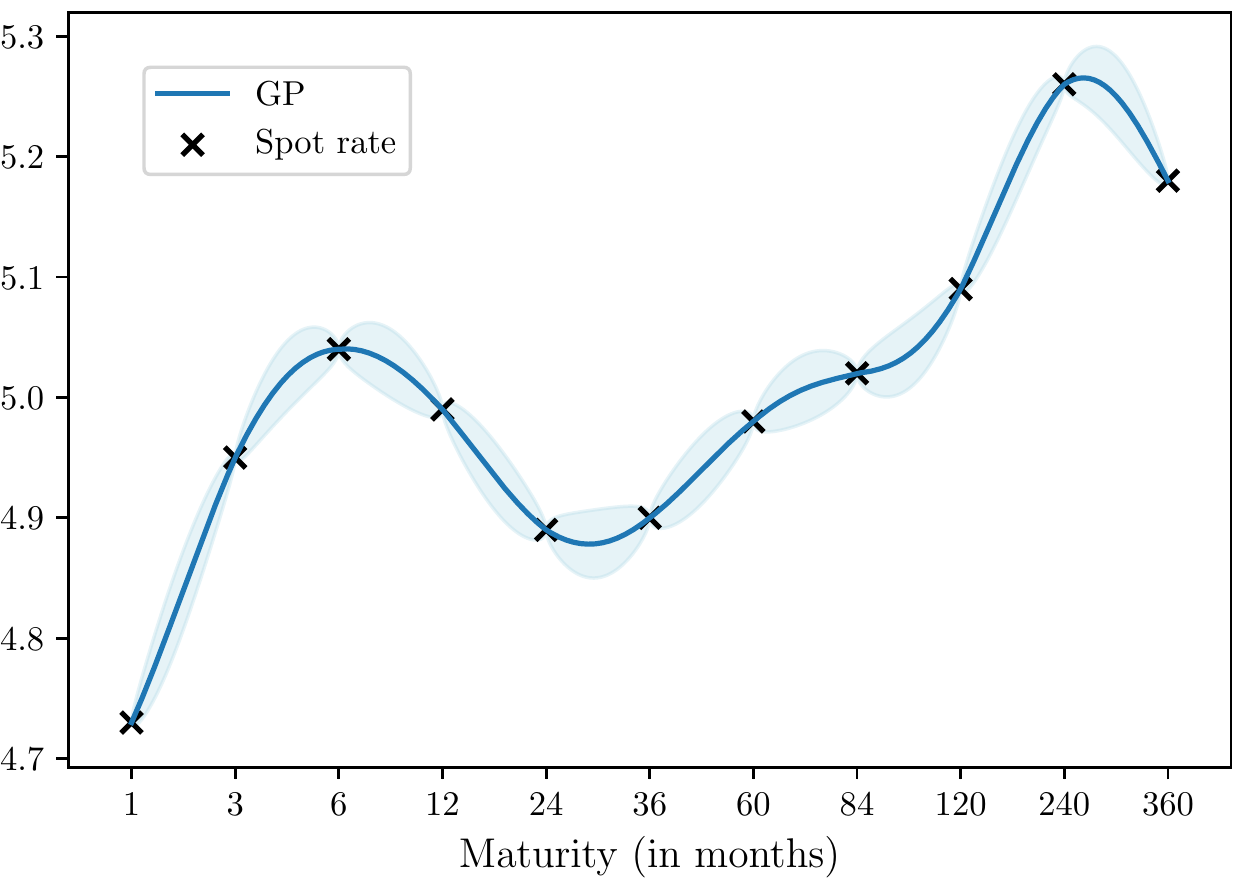}
\end{figure}

\begin{figure}[tbph]
\centering
\caption{GP fitting of the yield curve (June 2012)}
\label{fig:yieldcurve2}
\figureskip
\includegraphics[width = \figurewidth, height = \figureheight]{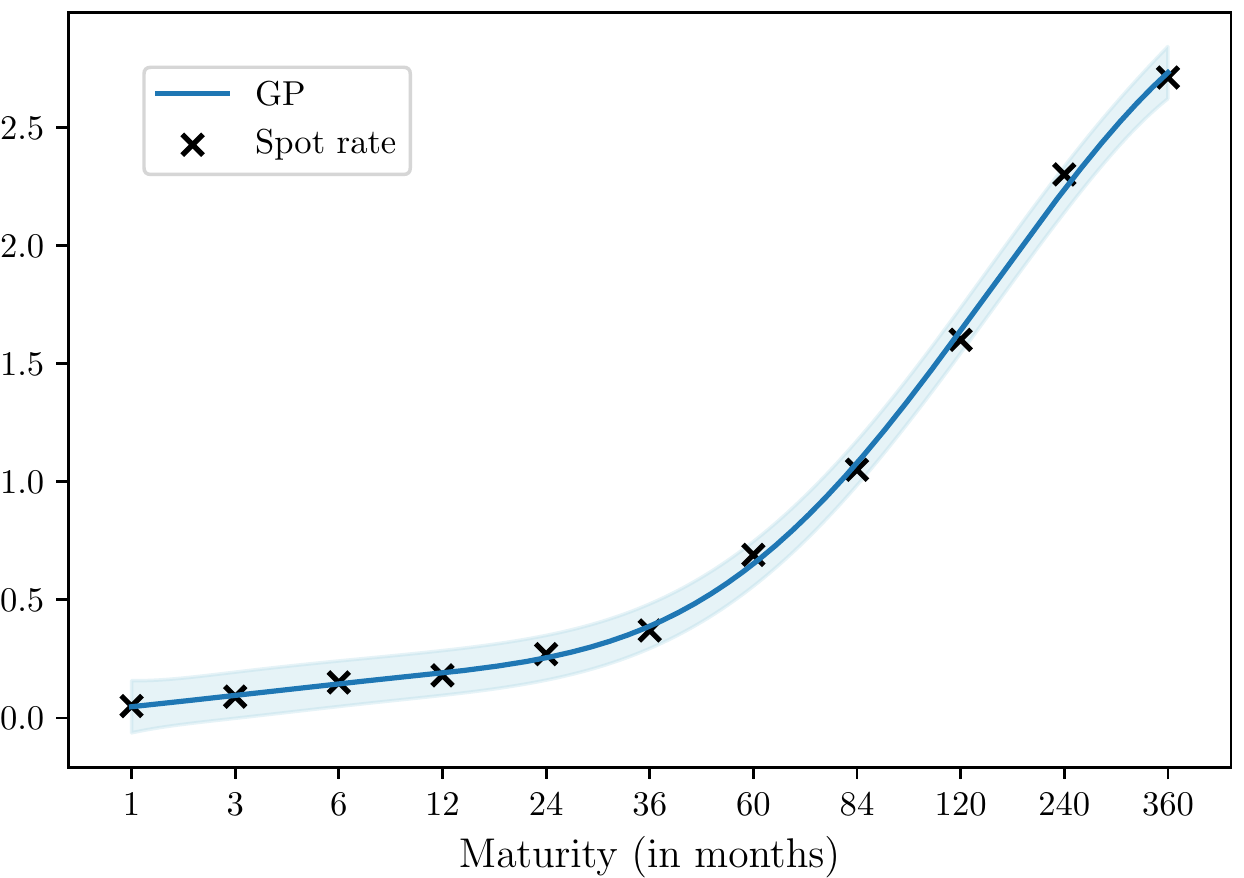}
\end{figure}

To illustrate the potential of GP methods in finance, we first consider the
fitting of the U.S. yield curve. One of the most used models is the
Nelson-Siegel parametric model, for which problems have been reported regarding
parameter estimation and their large variations over time (Annaert \textsl{et
al.}, 2013). GP can be thought of as a Bayesian nonparametric alternative. We
display in Figures \ref{fig:yieldcurve1} and \ref{fig:yieldcurve2} the fitting
of the yield curve corresponding to two different dates and presenting
different shapes\footnote{The covariance kernel is
$\mathcal{K}_{\mathrm{SE}}\cdot
\mathcal{K}_{\mathrm{EXP}}+\mathcal{K}_{\mathrm{RQ}}$, where the exponential
kernel $\mathcal{K}_{\mathrm{EXP}}$ is equal to:
\begin{equation*}
\mathcal{K}_{\mathrm{EXP}}\left( x,x^{\prime }\right) =\sigma ^{2}\exp
\left( -\frac{\left\Vert x-x^{\prime }\right\Vert _{2}}{2\ell }\right)
\end{equation*}}.

We now try GP-based methods to forecast movements of the U.S. yield curve. It
is a classical macroeconomic factor that can be used as a signal in the
forecasting of equity and bond returns (Rebonato, 2015; Cochrane \textsl{et
al.} (2005)) and thus is of practical interest in quantitative asset management.
Several approaches exist in time-series prediction with Gaussian processes. In
this paper, we mainly focus on the GP-ARX model, which is an application of ARX
models in the GP framework. The ARX model assumes a nonlinear relationship
between the time-series $Y_{t}$ and its previous values plus some exogenous
factors $X_{t}$:
\begin{equation*}
Y_{t}=f\left( Y_{t-1},Y_{t-2},\dots ,X_{t-1},X_{t-2},\dots \right)
+\varepsilon _{t}
\end{equation*}%
where $\varepsilon _{t}\sim \mathcal{N}\left( 0,\sigma _{\varepsilon
}^{2}\right) $ is a white noise process. The main idea of GP-ARX is to use a
Gaussian process surrogate for the function $f\sim \mathcal{GP}\left(
\mathbf{0},\mathcal{K}\right) $. Once the kernel $\mathcal{K}$ is chosen, the
training set simply consists in observations of $X_{t}$ and $Y_{t}$. Inference
of hyperparameters is done with the method of maximum likelihood as explained
in the previous section. Chandorkar \textsl{et al.} (2017) use this model to
forecast weather data and note the importance of the \textquotedblleft
\textit{persistence prediction}\textquotedblright\ in model building.
Persistence prediction is taking the current value as a forecast for tomorrow:
$\hat{Y}_{t+1}=Y_{t}$. The persistence model assumes then that the process is a random walk.
For time-series which exhibit memory effects, this
trivial forecast yields very good results if we use usual accuracy measures,
such as mean squared error. One way to measure memory in time-series is to
compute the Hurst exponent, which measures the long-term autocorrelation. A
Hurst exponent $H>\frac{1}{2}$ indicates positive long-term autocorrelation
while $H<\frac{1}{2}$ is the opposite\footnote{A Brownian motion has a Hurst
exponent of exactly $\frac{1}{2}$ and is memory-less.}. The Hurst exponent is
related to the fractal dimension of the time-series, and can be an indicator of
the predictability of the time-series (Kroha and \v{S}koula, 2018).\smallskip

\begin{table}[tbph]
\centering
\caption{Hurst exponent of U.S. spot rates}
\label{tab:yieldcurve1}
\begin{tabular}{cccccccccc}
\hline
Maturity       & 1M     & 3M     & 6M     & 1Y     & 2Y     & 5Y     & 10Y    & 20Y    & 30Y    \\
Hurst          & $0.40$ & $0.50$ & $0.61$ & $0.62$ & $0.57$ & $0.51$ & $0.49$ & $0.48$ & $0.50$ \\
\hline
\end{tabular}
\end{table}

\begin{figure}[tbph]
\centering
\caption{Prediction of the 2-year spot rate}
\label{fig:yieldcurve4}
\figureskip
\includegraphics[width = \figurewidth, height = \figureheight]{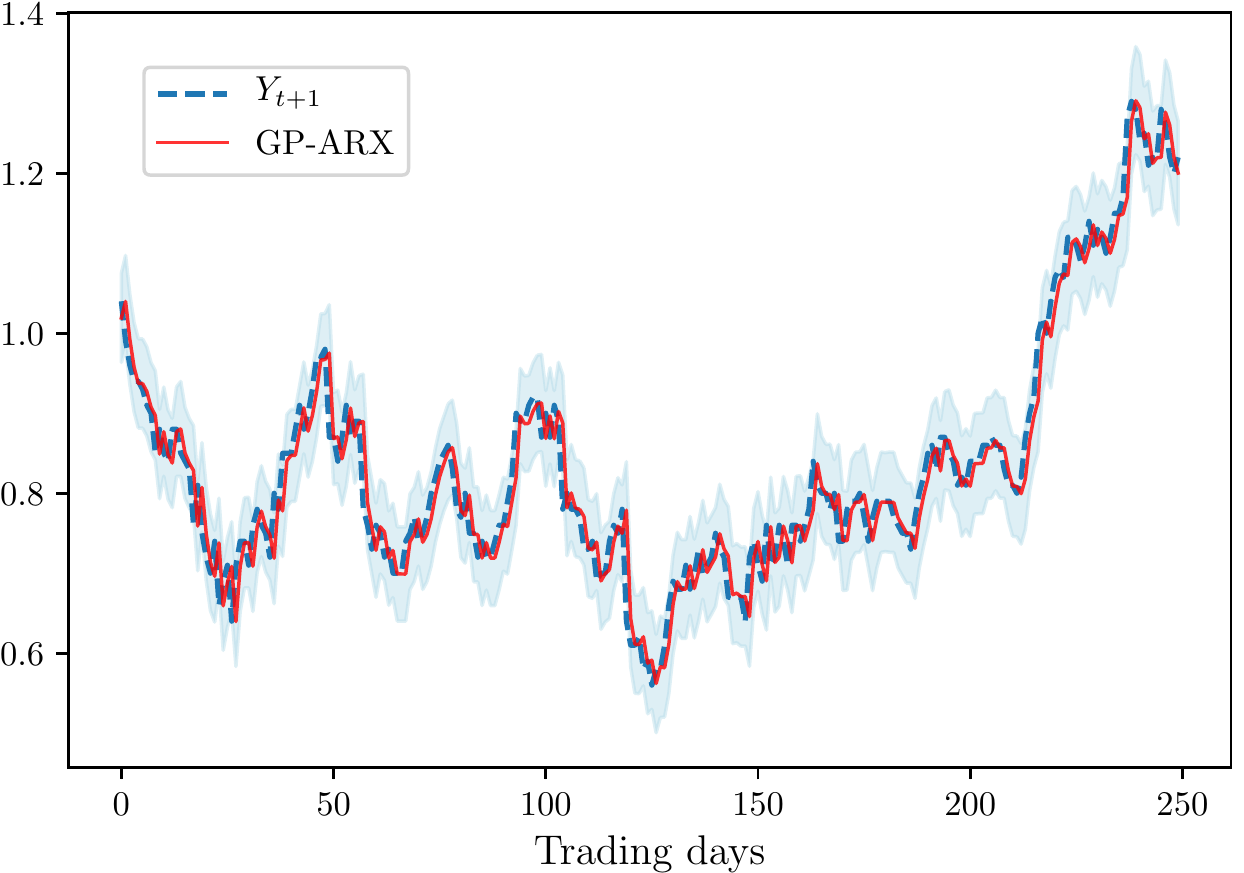}
\end{figure}

\begin{figure}[tbph]
\centering
\caption{Prediction of the 10-year spot rate}
\label{fig:yieldcurve5}
\figureskip
\includegraphics[width = \figurewidth, height = \figureheight]{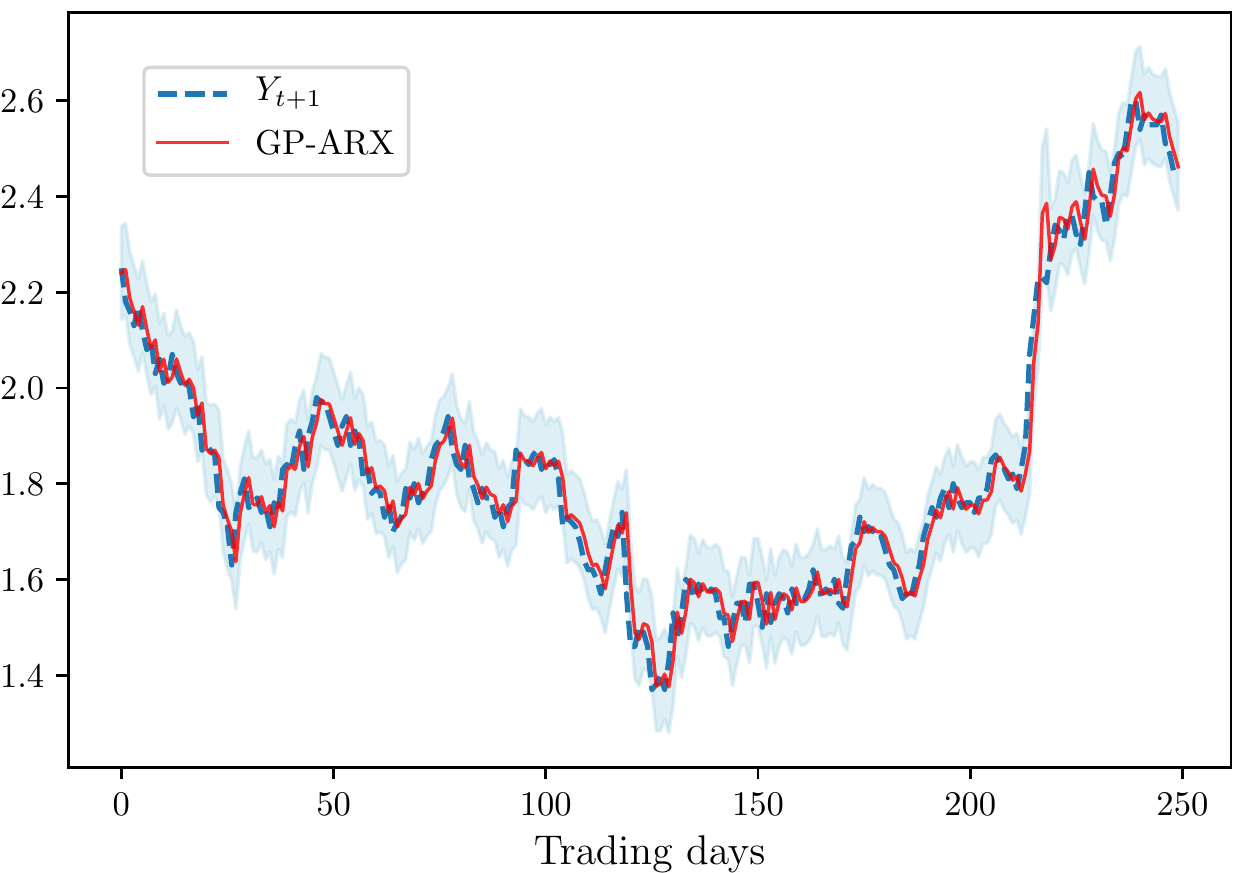}
\end{figure}

Our dataset consists in daily zero-coupon yield for the following maturities:
1, 3 and 6 months, 1, 2, 5, 10, 20 and 30 years. Table \ref{tab:yieldcurve1}
shows the Hurst exponent for the spot rates of different maturities. In Figures
\ref{fig:yieldcurve4} and \ref{fig:yieldcurve5}, we report the one-day ahead
rolling prediction during 2016, and the $95\%$ confidence interval. We
also forecast the 2Y and 10Y spot rates by using for each maturity a lag of
one for the GP-ARX model, and the three-month spot rate and its first lag for
the exogenous variable. Both spot rates exhibit persistence as it can be seen in
Table \ref{tab:yieldcurve2}. We notice that the three methods are equivalent in
this simplistic example. However, the GP-ARX method is able to estimate the
confidence interval, and this metric can be used as a trading signal.\smallskip

\begin{table}[htbp]
\centering
\caption{Root mean squared error (in \%)}
\label{tab:yieldcurve2}
\begin{tabular}{cccc}
\hline
Spot Rate & Persistence & ARX  & GP-ARX \\
\hline
{\TsV}2Y  & $3.33$           & $3.33$ & $\mathbf{3.32}$ \\
     10Y  & $\mathbf{4.40}$  & $4.53$ &          $4.45$ \\
\hline
\end{tabular}
\end{table}

\begin{remark}
The choice of the kernel function and the features are primordial.
The ARX model is actually equivalent to the GP-ARX model, when the kernel is linear.
In our GP-ARX model, we used a sum of exponential and linear
kernels and only used simple time-series features. Thanks to the ability to
capture different length scales and patterns with the choice of the kernel, the GP approach
can be used as good interpolators for a wide range of financial
applications by finding the suitable kernel combination.
\end{remark}

There is a growing interest in using Student-$t$ distribution instead of
Gaussian distribution in GP regression (Shah \textsl{et al.}, 2014). More
specifically, Chen \textsl{et al.} (2014) introduce a framework for multivariate
Gaussian and Student-$t$ process regression, and use it for stock and equity
index predictions. The problem of yield curve forecasting is intrinsically
multivariate. We are trying to predict a vector of interest rates for different
maturities, that are highly correlated and dependent. Instead of treating each
output separately, Student-$t$ multivariate process regression works with a
dataset consisting of full yield curve observations as inputs and outputs at
the same time, while taking into account output correlations. We recall in
Appendix \ref{appendix:mvt} on page \pageref{appendix:mvt} the definition and
some useful properties of multivariate and matrix-variate Student-$t$
distributions. Specifically, as for the Gaussian case, the posterior distribution
is still a matrix-variate Student-$t$ distribution, which makes computations of
inference step and maximum marginal likelihood tractable.
This motivates the definition of the Student-$t$ processes (TP). A
collection of random vectors is a TP if and only if any finite number of them
has a joint multivariate Student-$t$ distribution. We have tested TP-ARX in
place of GP-ARX. Unfortunately, we have not found better forecasting values.
This result is disappointing since we may think that one of the issues in yield
curve modeling is the cross-section correlation of spot rates%
\footnote{TPs are especially designed to take into account both
cross-section and time-series correlations, whereas GPs can only consider the
dynamics of one direction (cross-section or time-series), not both
(See Equations (7) and (15) in Chen \textsl{et al} (2018)).} and the
possible fat tails that we observe in fixed-income assets.

\subsection{Portfolio optimization}

We now consider an application of Bayesian optimization to portfolio
optimization in the context of quantitative asset management. We first describe
the asset allocation problem, which is a trend-following strategy, then we show
how to solve it and finally we use Bayesian optimization with a squared exponential kernel
to find optimal hyperparameters of the trend-following strategy.

\subsubsection{The trend-following strategy}

We consider a universe of $n$ assets for which we observe daily prices and
we look for an optimal portfolio, that is an allocation vector $x\in \mathbb{%
R}^{n}$ that balances risk and return. If we can predict the vector $\mu $ of
expected returns and compute the covariance matrix $\Sigma $ of asset returns,
then the regularized Markowitz optimization problem (Roncalli, 2013; Bourgeron
\textsl{et al.}, 2018) is the following:
\begin{equation*}
x^{\star }\left( \gamma\right) =\arg \min_{x}\frac{1}{2}x^{\top
}\Sigma x-\gamma \mu ^{\top }x+\lambda \left\Vert x-x_{0}\right\Vert _{2}^{2}
\end{equation*}%
where $\gamma $ is the inverse of the risk-aversion coefficient, $\lambda $ is the ridge
regularization parameter and $x_{0}$ is a reference portfolio. We consider a
simple version of the trend-following strategy:
\begin{itemize}
\item The expected returns are computed using a moving-average estimator. Let
    $P_{i,t}$ be the daily price of Asset $i$. We have:
    \begin{equation*}
        \mu _{i,t}=\frac{P_{i,t}}{P_{i,t-\ell \left( \mu \right) }}-1
    \end{equation*}%
    where $\ell \left( \mu \right) $ is the window length of the MA
    estimator.
\item The covariance matrix is estimated using the empirical estimator, the
    window length of which is denoted by $\ell \left( \Sigma \right) $.
\item The portfolio is rebalanced at fixed dates $t$, for example on a
    monthly or weekly basis.
\end{itemize}
Let $x_{t}$ be the optimal portfolio at the rebalancing date $t$. The
allocation problem is given by:
\begin{eqnarray}
x_{t}\left( \lambda \right)  &=&\arg \min_{x}-\mu _{t}^{\top }x+\lambda
\left\Vert x-x_{t-1}\right\Vert _{2}^{2}  \label{eq:trend1} \\
&\text{s.t.}&\sigma _{t}\left( x\right) \leq \bar{\sigma}  \notag
\end{eqnarray}%
where $\mu _{t}$ is the estimated vector of expected returns at time $t$, $%
\sigma _{t}\left( x\right) =\sqrt{x^{\top }\Sigma _{t}x}$ is the portfolio
volatility estimated at time $t$, $\bar{\sigma}$ is the target volatility of
the trend-following strategy. To solve this convex problem, we use the ADMM
algorithm given in Appendix \ref{appendix:admm} on page \pageref{appendix:admm}
(Boyd \textsl{et al.}, 2011). Following Bourgeron \textsl{et al.} (2018) and
Richard and Roncalli (2019), it is natural to write the previous problem as
follows:
\begin{eqnarray*}
x_{t} &=&\arg \min_{x}-\mu _{t}^{\top }x+\lambda \left\Vert
x-x_{t-1}\right\Vert _{2}^{2}+\mathds{1}_{\Omega }\left( z\right)  \\
&\text{s.t.}&x-z=0
\end{eqnarray*}%
where $\Omega =\left\{ z\in \mathbb{R}^{n}:\left\Vert z^{\top }\Sigma
_{t}z\right\Vert _{2}^{2}\leq \bar{\sigma}^{2}\right\} $. However, we
improve the ADMM algorithm by introducing the Cholesky trick:
\begin{eqnarray*}
x_{t} &=&\arg \min_{x}-\mu _{t}^{\top }x+\lambda \left\Vert
x-x_{t-1}\right\Vert _{2}^{2}+\mathds{1}_{\Omega }\left( z\right)  \\
&\text{s.t.}&-L_{t}x+z=0
\end{eqnarray*}%
where $\Omega =\left\{ z\in \mathbb{R}^{n}:\left\Vert z\right\Vert _{2}^{2}\leq
\bar{\sigma}^{2}\right\} $ and $L_{t}$ is the upper Cholesky decomposition
matrix of $\Sigma _{t}$. It follows that $z=L_{t}x$ and:
\begin{eqnarray*}
\left\Vert z\right\Vert _{2}^{2} &=&z^{\top }z \\
&=&x^{\top }L_{t}^{\top }L_{t}x \\
&=&x^{\top }\Sigma _{t}x \\
&=&\sigma _{t}^{2}\left( x\right)
\end{eqnarray*}%
In fact, our experience shows that the Cholesky trick helps to accelerate
the convergence of the ADMM algorithm with respect to the formulation of
Bourgeron \textsl{et al.} (2018) and Richard and Roncalli (2019). Finally,
it follows that the ADMM algorithm becomes:
\begin{eqnarray*}
x^{(k+1)} &=&\arg \min -\mu _{t}^{\top }x+\lambda \left\Vert
x-x_{t-1}\right\Vert _{2}^{2}+\frac{\varphi }{2}\left\Vert
-L_{t}x+z^{(k)}+u^{(k)}\right\Vert _{2}^{2} \\
z^{(k+1)} &=&\arg \min \mathds{1}_{\Omega }\left( z\right) +\frac{\varphi }{2%
}\left\Vert -L_{t}x^{(k+1)}+z+u^{(k)}\right\Vert _{2}^{2} \\
u^{(k+1)} &=&u^{(k)}-L_{t}x^{(k+1)}+z^{(k+1)}
\end{eqnarray*}%
We notice that the $z$-step corresponds to a simple projection on the
Euclidean ball of center $0$ and radius $\bar{\sigma}$ of the vector $%
L_{t}x^{(k+1)}-u^{(k)}$, and the computation of the proximal operator is
straightforward. The $x$-step corresponds to a linear system. If we define $%
f^{\left( k\right) }\left( x\right) $ as follows:
\begin{equation*}
f^{\left( k\right) }\left( x\right) =-\mu _{t}^{\top }x+\lambda \left\Vert
x-x_{t-1}\right\Vert _{2}^{2}+\frac{\varphi }{2}\left\Vert
-L_{t}x+z^{(k)}+u^{(k)}\right\Vert _{2}^{2}
\end{equation*}%
we deduce that:%
\begin{eqnarray*}
\nabla f^{\left( k\right) }\left( x\right)  &=&-\mu _{t}+\lambda x-\lambda
x_{t-1}+\varphi L_{t}^{\top }L_{t}x-\varphi L_{t}^{\top }\left(
z^{(k)}+u^{(k)}\right)  \\
&=&-\mu _{t}+\lambda x-\lambda x_{t-1}+\varphi \Sigma _{t}x-\varphi
L_{t}^{\top }\left( z^{(k)}+u^{(k)}\right)
\end{eqnarray*}%
Finally, we obtain the following solution:
\begin{equation*}
x^{(k+1)}=\left( \varphi \Sigma _{t}+\lambda I_{n}\right) ^{-1}\left( \mu
_{t}+\lambda x_{t-1}+\varphi L_{t}^{\top }\left( z^{(k)}+u^{(k)}\right)
\right)
\end{equation*}

\subsubsection{Hyperparameter estimation of the trend-following strategy}

The trend-following strategy depends on three hyperparameters:
\begin{enumerate}
\item the parameter $\lambda $ that controls the turnover between two
    rebalancing dates;

\item the window length $\ell \left( \mu \right) $ that controls the
    estimation of trends;

\item the horizon time $\ell \left( \Sigma \right) $ that measures the risk
    of the assets.
\end{enumerate}
Traditionally, the trend-following strategy is implemented by considering that
these hyperparameters are fixed. By construction, their choice has a big impact
on the strategy design. For instance, a small value of $\ell \left( \mu \right)
$ will catch short momentum, whereas a large value of $\ell \left( \mu \right) $
will look for more persistent trends. This hyperparameter is then key for
distinguishing short-term and long-term CTAs.\smallskip

In fact, the right asset allocation problem is not given by Equation
(\ref{eq:trend1}), but is defined as follows:
\begin{eqnarray}
x_{t}\left( \lambda _{t},\ell _{t}\left( \mu \right) ,\ell _{t}\left( \Sigma
\right) \right)  &=&\arg \min_{x}-\mu _{t}^{\top }x+\lambda _{t}\left\Vert
x-x_{t-1}\right\Vert _{2}^{2}  \label{eq:trend2} \\
&\text{s.t.}&\sigma _{t}\left( x\right) \leq \bar{\sigma}  \notag
\end{eqnarray}%
This means that the hyperparameters are not fixed and must be estimated at each
rebalancing date. In the previous framework, the estimation consists in
finding $x_{t}$ given that $\lambda $, $\ell \left( \mu \right) $ and $\ell
\left( \Sigma \right) $ are constant. In our framework, the estimation consists
in finding the optimal portfolio $x_{t}$, but also the optimal parameters
$\lambda _{t}$, $\ell _{t}\left( \mu \right) $ and $\ell _{t}\left( \Sigma
\right) $. This can be done using the Bayesian optimization
framework.\smallskip

By nature, the parameters $\ell \left( \mu \right) $ and $\ell \left( \Sigma
\right) $ are discrete and generally expressed in months, e.g. $\ell \left( \mu
\right) \in \left\{ 3,6,12,24\right\} $ and $\ell \left( \Sigma \right) \in
\left\{ 3,6,12\right\} $. Discrete, integer or categorical parameters are not
easy to manage in Bayesian optimization since Gaussian processes (or random
forests), which serve as surrogates for the black-box function, are not
adapted. Since no standard approach exists yet\footnote{Although some new
methods are emerging (Garrido-Merch\'an and Hern\'andez-Lobato, 2017).}, we use
a simple method which consists in using continuous variables in the Bayesian
optimization step while flooring the hyperparameters $\ell \left( \mu \right) $
and $\ell \left( \Sigma \right) $ to the nearest integer when computing the
objective function given by Equation (\ref{eq:trend2}).\smallskip

The choice of the objective function is the main step when implementing
Bayesian optimization. In a classical machine learning problem, the objective
function can be the cross-validation score of the hyperparameters, in order to
reduce the risk of overfitting. Defining an objective for a quantitative
strategy is less clear and prone to overfitting. One simple and obvious
function is the Sharpe ratio of the strategy. Each rebalancing date, we run a
Bayesian optimization to look for hyperparameters which maximizes the
historical Sharpe ratio over a given period. A more robust objective function
is the minimum of the rolling Sharpe ratio in order to reduce the overfitting
bias. For example, we can compute the rolling six-month Sharpe ratio
$\limfunc{SR}_{\tau }\left( \lambda ,\ell \left( \mu \right) ,\ell \left(
\Sigma \right) \right) $ for a backtest with fixed hyperparameters and a period
$\left[ \tau -0.5;\tau \right] $ and use Bayesian optimization to solve:
\begin{equation*}
\left\{ \lambda _{t},\ell _{t}\left( \mu \right) ,\ell _{t}\left( \Sigma
\right) \right\} =\arg \max \left\{ \min_{\tau \in \left[ t-2,t\right[ }%
\limfunc{SR}\nolimits_{\tau }\left( \lambda ,\ell \left( \mu \right) ,\ell
\left( \Sigma \right) \right) \right\}
\end{equation*}%
However, in order to benefit from the exploration of the parameters space by
Bayesian optimization, we use another approach. Since volatility is already
controlled via the portfolio optimization constraint and regularization, we
prefer to choose the return of the strategy over a two-year historical
period for the objective function:%
\begin{equation*}
\left\{ \lambda _{t},\ell _{t}\left( \mu \right) ,\ell _{t}\left( \Sigma
\right) \right\} =\arg \max \hat{\mu}_{t}\left( \lambda ,\ell \left( \mu
\right) ,\ell \left( \Sigma \right) \right)
\end{equation*}%
where $\hat{\mu}_{t}\left( \lambda ,\ell \left( \mu \right) ,\ell \left(
\Sigma \right) \right) $ is the performance of the backtest for the period $%
\left[ t-2;t\right] $. We keep track of all samples tested during Bayesian
optimization, sort them according to their objective function and select the
best three sets of hyperparameters to compute three different optimal weights
that are averaged to form the final portfolio. This approach
considerably reduces the overfitting bias.

\subsubsection{An example}

Our dataset consists of daily prices of 13 futures contracts on world-wide
equity indices such as the S\&P 500 and Eurostoxx indices and 10Y sovereign bonds
from 2006 to 2017. The Bayesian optimization strategy described in the previous
paragraph is compared to the basic one-year trend-following strategy where
parameters remain unchanged through time: $\lambda _{t}$ is set to $10\%$ while
$\ell _{t}\left( \mu \right) $ and $\ell _{t}\left( \Sigma \right) $ are equal
to $12$ months.\smallskip

\begin{table}[tbph]
\centering
\caption{Backtest results (2006 -- 2016)}
\label{tab:trend1}
\begin{tabular}{ccccc}
\hline
\multirow{2}{*}{Strategy} & \multirow{2}{*}{Sharpe ratio} &  \multirow{2}{*}{Return} & \multirow{2}{*}{Volatility} & \multirow{2}{*}{MDD} \\
         &              &         &            &          \\ \hline
Naive    &       $1.56$ & $5.9\%$ &    $3.7\%$ & -$5.7\%$ \\
Bayesian &       $1.71$ & $7.4\%$ &    $4.3\%$ & -$4.7\%$ \\
\hline
\end{tabular}
\end{table}

\begin{figure}[tbph]
\centering
\caption{Cumulative performance of trend-following strategies}
\label{fig:trend-perf}
\figureskip
\includegraphics[width = \figurewidth, height = \figureheight]{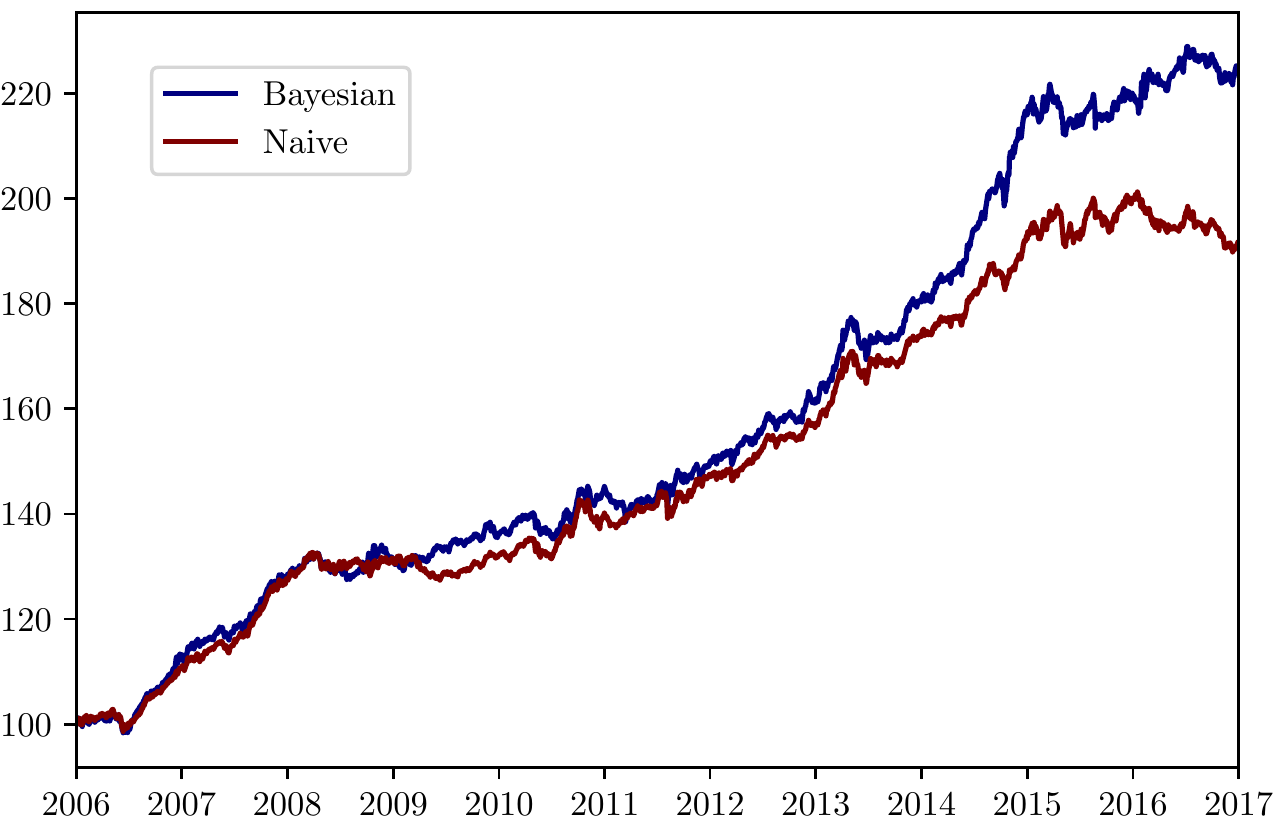}
\end{figure}

The cumulative performance of the strategies is shown in Figure \ref{fig:trend-perf},
whereas Table \ref{tab:trend1} shows the results of the two
strategies. We notice that Bayesian optimization is able to improve the
annualized Sharpe ratio from $1.56$ to $1.71$ and reduce drawdowns. However, we
do not believe that this result is important, because it is just a backtest.
More interesting are the dynamics of the hyperparameters estimated by the
Bayesian optimization. On page \pageref{fig:trend-lambda}, we report the
dynamics of $\lambda _{t}$ (Figure \ref{fig:trend-lambda}), $\ell _{t}\left(
\mu \right) $ (Figure \ref{fig:trend-mu}) and $\ell _{t}\left( \Sigma \right) $
(Figure \ref{fig:trend-cov}), whereas the statistics are reported in Table
\ref{tab:trend2}. We observe that $\lambda _{t}$ moves relatively fast. At the
beginning of the 2008 Global Financial Crisis, it is reduced implying a more
reactive allocation. At the end of the 2008 crisis, we observe the opposite
effect. $\lambda _{t}$ is increased and the allocation becomes less reactive.
However, this must be compared with the dynamics of $\ell _{t}\left( \mu
\right) $ and $\ell _{t}\left( \Sigma \right) $. Most of the time, the optimal
window $\ell _{t}\left( \mu \right) $ is high and is equal to 18 months on
average. However, during and after the GFC, $\ell _{t}\left( \mu \right) $ is
dramatically reduced. The parallel can be done with the performance of
short-term and long-term CTAs. On average, long-term CTAs outperform short-term
CTAs, but during some periods, short-term CTAs can do a very good job, and post
an incredible performance while long-term CTAs have a strong negative
performance. Concerning $\ell _{t}\left( \Sigma \right) $, the results indicate
that a short-term window is better, while the market practice is to consider
long-term window (typically a one-year empirical covariance matrix). In fact,
there is a trade-off between the regularization parameter $\lambda _{t} $ and
the covariance window $\ell _{t}\left( \Sigma \right) $. Our model chooses a
short-term covariance, because it can control the turnover thanks to the ridge
parameter.\smallskip

\begin{table}[tbph]
\centering
\caption{Statistics of optimal hyperparameters (2006 -- 2016)}
\label{tab:trend2}
\begin{tabular}{cccccc}
\hline
\multirow{2}{*}{$\theta _{t}$} & \multirow{2}{*}{$\min \theta _{t}$} & \multirow{2}{*}{$\max \theta _{t}$} &
\multirow{2}{*}{$\bar{\theta}_{t}$} & \multirow{2}{*}{$\sigma \left( \theta _{t}\right) $} &
\multirow{2}{*}{$\rho \left( \theta _{t},\limfunc{VIX}_{t}\right)$} \\
                                     &        &           &              &              &                 \\ \hline
$\lambda _{t}$                       & $0.01$ & ${\TsV}2$ & ${\TsV}0.13$ &      $0.056$ & ${\TsVIII}48\%$ \\
$\ell _{t}\left( \mu \right)$        & $3.00$ &      $24$ &      $17.45$ & $8.10{\TsV}$ &         $-49\%$ \\
$\ell _{t}\left( \Sigma \right)$     & $3.00$ &      $12$ & ${\TsV}4.13$ & $1.72{\TsV}$ & ${\TsVIII}38\%$ \\
\hline
\end{tabular}
\end{table}

\begin{figure}[tbph]
\centering
\caption{Comparison of $\ell_{t}\left( \mu \right) $ and $\func{VIX}_{t}$}
\label{fig:trend-vix}
\figureskip
\includegraphics[width = \figurewidth, height = \figureheight]{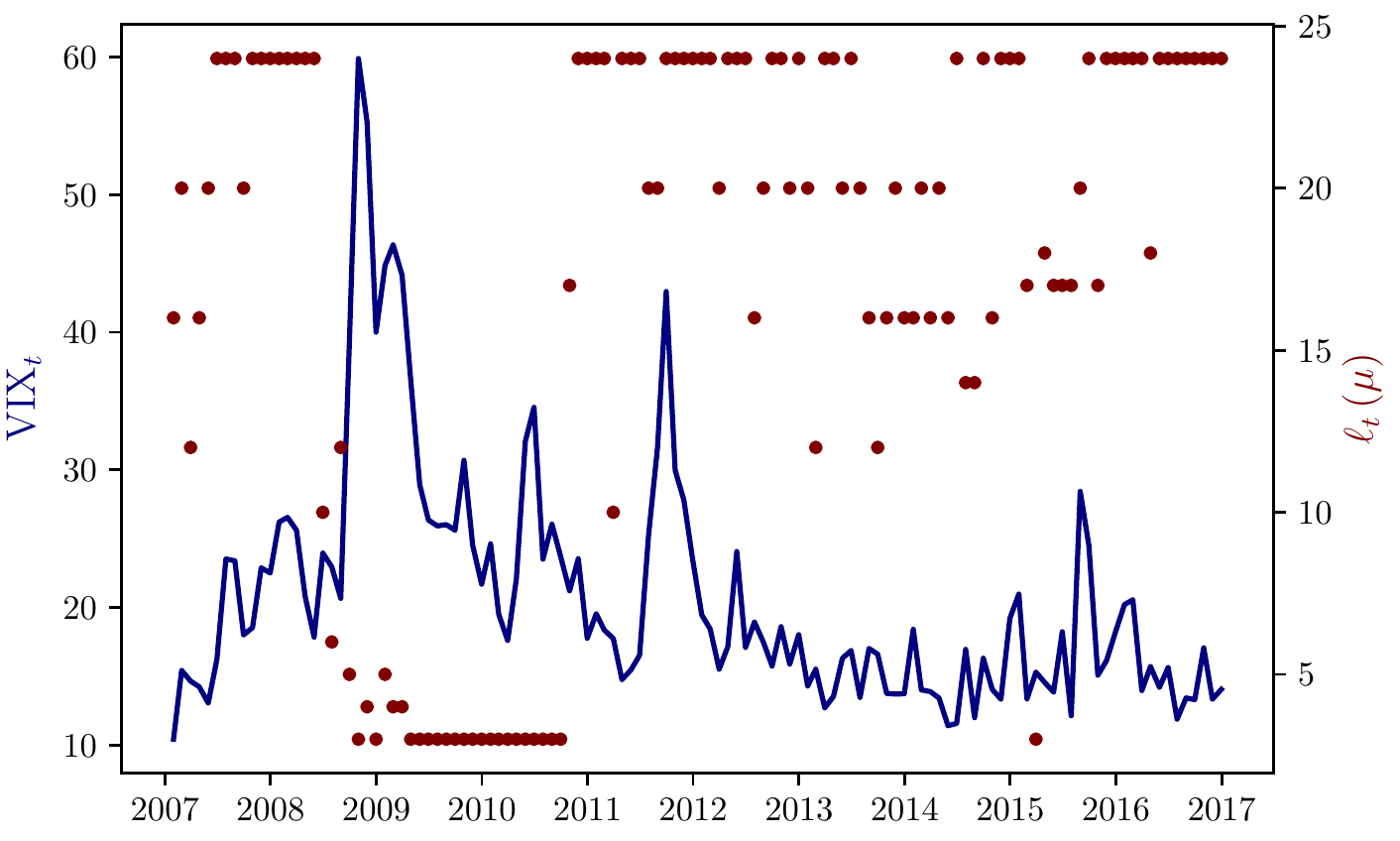}
\end{figure}

The dynamics of these parameters can be analyzed with respect to market
volatility. For instance, if we compute the correlation with the VIX index, we
observe that the regularization hyperparameter $\lambda _{t}$ and the
covariance window $\ell _{t}\left( \Sigma \right) $ show a positive correlation
with $\limfunc{VIX}_{t}$ while the correlation is negative between $\ell
_{t}\left( \mu \right) $ and $\limfunc{VIX}_{t}$ (see Table \ref{tab:trend2}).
This indicates that the strategy focuses on short-term momentum and takes less
risk in times of high volatility as it is shown in Figure \ref{fig:trend-vix}
during 2008. The hyperparameter $\lambda _{t}$, which expresses a turnover
penalty, then has a conservative effect during the 2008 Global Financial
Crisis.

\begin{remark}
Because of local minima (since the objective function might not be well-behaved
or even regular when considering categorical variables), periods of
instability occur in the estimation of the optimal parameters. This is the
case in 2013-2014 (Figure \ref{fig:trend-vix}). In such cases, we benefit from
diversification when mixing the three optimal portfolios unlike when selecting only
one solution.
\end{remark}

\section{Conclusion}

In this paper, we explore the use of Gaussian processes and Bayesian
optimization in finance. Two applications have been considered: the yield curve
modeling and the online calibration of trend-following strategies. Our
results show that GPs are a powerful tool for fitting the yield curve. Therefore,
GPs can be used as a semi-parametric alternative of popular parametric
approaches such as the Nelson-Siegel model. However, our results also show that
GPs are equivalent to traditional econometric approaches for forecasting
interest rates, but they do not do a better job. The case of trend-following strategies is more interesting,
because it is a classic problem in finance when we choose ex-ante the value of
hyperparameters. Until now, there was no other alternative approach to test
several combinations of hyperparameters, and to choose the best combination
in trying to avoid the in-sample bias, which is inherent to any backtesting
protocol. We show how to implement a Bayesian optimization for estimating the
window lengths of the trend vector and the covariance matrix.
Results confirm the practice and what we observe in the industry of CTA and
dynamic risk parity funds. Generally, it is better to consider a long window for the
expected returns and a short window for the risks. However, there is a
trade-off between performance, turnover and rebalancing costs. This is why it
is necessary to introduce penalty functions in the portfolio optimization.
Our results also show that there are some periods where reducing the window of trends may
add value. In this context, the Bayesian optimization provides a normative way
to build online trend following strategies.

\clearpage

\appendix

\section{Mathematical results}

\subsection{Notations}
\label{appendix:notation}

We use the following notations:
\begin{itemize}

\item $I_n$ is the identity matrix of $\mathbb{R}^n$.

\item $\mathbf{1}$ is a vector of ones.

\item $\mathds{1}_{\Omega}\left( x\right) $ is the convex indicator function
    of $\Omega$: $ \mathds{1}_{\Omega}\left( x\right) =0$ for $x\in \Omega$
    and $\mathds{1}_{\Omega}\left( x\right) = +\infty $ for $x\notin \Omega$.
\item $x^+$ is the positive part $\max\left(0, x\right)$ of $x$.

\item $\Phi\left(x\right)$ is the standard normal cumulative distribution
    function:
    \begin{equation*}
        \Phi\left(x\right) = \frac{1}{\sqrt{2\pi}} \int_{-\infty}^{x} e^{-\frac{1}{2} t^2} \mathrm{d}t
    \end{equation*}
    whereas $\phi\left(x\right)$ is its probability density function:
    \begin{equation*}
        \phi\left(x\right) = \frac{1}{\sqrt{2\pi}}e^{-\dfrac{1}{2} x^2}
    \end{equation*}

\item $\Gamma\left(s\right)$ is Euler's gamma function:
    \begin{equation*}
        \Gamma\left(s\right) = \int_{0}^{\infty} x^{s-1} e^{-x} \mathrm{d}x
    \end{equation*}

\item $\Gamma_n$ is the multivariate gamma function:
    \begin{equation*}
        \Gamma_n\left(s\right) = \pi^{\dfrac{n\left(n-1\right)}{2}} \prod_{i=1}^{n} \Gamma\left(s + \frac{\left(1-i\right)}{2}\right)
    \end{equation*}

\item $f\left( x\right) \sim \mathcal{GP}\left( m\left( x\right) ,\mathcal{K}
    \left( x,x\right) \right) $ denotes a Gaussian process.

\item $\hat{f}\left( x^{\star }\right) =f\left( x^{\star }\mid x,y\right) $
    is the random vector of outputs conditional to the sample $\left(
    x,y\right) $.

\item $\hat{m}\left( x^{\star }\right) =m\left( x^{\star }\mid x,y\right) $
    is the conditional expectation of $x^{\star }$ with respect to the sample
    $\left( x,y\right) $.

\item $\mathcal{\hat{K}}\left( x^{\star },x^{\star }\right) =\mathcal{K}
    \left( x^{\star },x^{\star }\mid x,y\right) $ is the conditional
    covariance matrix of $x^{\star }$ with respect to the sample $\left(
    x,y\right) $.

\item $x=\left( x_{1},\ldots ,x_{n}\right) $ is a matrix of dimension $
    n\times d$.

\item $x^{\star }=\left( x_{1}^{\star },\ldots ,x_{n}^{\star }\right) $ is a
    matrix of dimension $n^{\star }\times d$.

\end{itemize}

\clearpage

\subsection{Conditional Gaussian distribution}
\label{appendix:section-conditional-expectation}

Let us consider a Gaussian random vector defined as follows:%
\begin{equation*}
\left(
\begin{array}{c}
X \\
Y%
\end{array}%
\right) \sim \mathcal{N}\left( \left(
\begin{array}{c}
\mu _{x} \\
\mu _{y}%
\end{array}%
\right) ,\left(
\begin{array}{cc}
\Sigma _{xx} & \Sigma _{xy} \\
\Sigma _{yx} & \Sigma _{yy}%
\end{array}%
\right) \right)
\end{equation*}%
Then, the marginal distributions of $X$ and $Y$ are given by $X\sim \mathcal{%
N}\left( \mu _{x},\Sigma _{xx}\right) $ and $Y\sim \mathcal{N}\left( \mu
_{y},\Sigma _{yy}\right) $, and we have $\func{cov}\left( X,Y\right) =\Sigma
_{xy}$. The conditional distribution of $Y$\ given $X=x$ is a multivariate
normal distribution:
\begin{equation*}
Y\mid X=x\sim \mathcal{N}\left( \mu _{y\mid x},\Sigma _{yy\mid x}\right)
\end{equation*}%
where:%
\begin{equation*}
\mu _{y\mid x}=\mathbb{E}\left[ Y\mid X=x\right] =\mu _{y}+\Sigma
_{yx}\Sigma _{xx}^{-1}\left( x-\mu _{x}\right)
\end{equation*}%
and:%
\begin{equation*}
\Sigma _{yy\mid x}=\sigma ^{2}\left[ Y\mid X=x\right] =\Sigma _{yy}-\Sigma
_{yx}\Sigma _{xx}^{-1}\Sigma _{xy}
\end{equation*}

\subsection{Derivation of the SoR approximation}
\label{appendix:section-sor-approximation}

The SoR approximation is based on the Woodbury matrix identity:%
\begin{equation}
\left( A+BCD\right) ^{-1}=A^{-1}-A^{-1}B\left( C^{-1}+DA^{-1}B\right)
^{-1}DA^{-1}  \label{eq:appendix-woodbury}
\end{equation}%
and four approximations:%
\begin{eqnarray}
\mathcal{K}\left( x,x\right)  &\approx &\mathcal{K}\left( x,x_{m}\right)
\mathcal{K}\left( x_{m},x_{m}\right) ^{-1}\mathcal{K}\left( x_{m},x\right)
\label{eq:appendix-approx1a} \\
\mathcal{K}\left( x^{\star },x\right)  &\approx &\mathcal{K}\left( x^{\star
},x_{m}\right) \mathcal{K}\left( x_{m},x_{m}\right) ^{-1}\mathcal{K}\left(
x_{m},x\right)   \label{eq:appendix-approx1b} \\
\mathcal{K}\left( x,x^{\star }\right)  &\approx &\mathcal{K}\left(
x,x_{m}\right) \mathcal{K}\left( x_{m},x_{m}\right) ^{-1}\mathcal{K}\left(
x_{m},x^{\star }\right)   \label{eq:appendix-approx1c} \\
\mathcal{K}\left( x^{\star },x^{\star }\right)  &\approx &\mathcal{K}\left(
x^{\star },x_{m}\right) \mathcal{K}\left( x_{m},x_{m}\right) ^{-1}\mathcal{K}%
\left( x_{m},x^{\star }\right)   \label{eq:appendix-approx1d}
\end{eqnarray}

\subsubsection{Preliminary results}

Using Equations (\ref{eq:appendix-woodbury}) and (\ref{eq:appendix-approx1a}%
), we deduce that:%
\begin{eqnarray}
(\ast ) &= &\left( \mathcal{K}\left( x,x\right) +\sigma_{\varepsilon} ^{2}I_{n}\right) ^{-1}
\notag \\
&\approx&\left( \sigma_{\varepsilon} ^{2}I_{n}+\mathcal{K}\left( x,x_{m}\right) \mathcal{K}%
\left( x_{m},x_{m}\right) ^{-1}\mathcal{K}\left( x_{m},x\right) \right) ^{-1}
\notag \\
&=&\frac{1}{\sigma_{\varepsilon} ^{2}}I_{n}-\frac{1}{\sigma_{\varepsilon} ^{2}}I_{n}\mathcal{K}\left(
x,x_{m}\right) \cdot   \notag \\
&&\left( \mathcal{K}\left( x_{m},x_{m}\right) +\mathcal{K}\left(
x_{m},x\right) \frac{1}{\sigma_{\varepsilon} ^{2}}I_{n}\mathcal{K}\left( x,x_{m}\right)
\right) ^{-1}\mathcal{K}\left( x_{m},x\right) \frac{1}{\sigma_{\varepsilon} ^{2}}I_{n}
\notag \\
&=&\frac{1}{\sigma_{\varepsilon} ^{2}}I_{n}-\frac{1}{\sigma_{\varepsilon} ^{2}}\mathcal{K}\left(
x,x_{m}\right) \left( \sigma_{\varepsilon} ^{2}\mathcal{K}\left( x_{m},x_{m}\right) +%
\mathcal{K}\left( x_{m},x\right) \mathcal{K}\left( x,x_{m}\right) \right)
^{-1}\mathcal{K}\left( x_{m},x\right)   \notag
\end{eqnarray}%
We obtain the following relationship:%
\begin{equation}
\left( \mathcal{K}\left( x,x\right) +\sigma_{\varepsilon} ^{2}I_{n}\right) ^{-1}\approx
\frac{1}{\sigma_{\varepsilon} ^{2}}I_{n}-\frac{1}{\sigma_{\varepsilon} ^{2}}\mathcal{K}\left(
x,x_{m}\right) \mathcal{\tilde{K}}\left( x_{m},x_{m}\right) ^{-1}\mathcal{K}%
\left( x_{m},x\right)   \label{eq:appendix-approx2}
\end{equation}%
where:%
\begin{equation*}
\mathcal{\tilde{K}}\left( x_{m},x_{m}\right) =\mathcal{K}\left(
x_{m},x\right) \mathcal{K}\left( x,x_{m}\right) +\sigma_{\varepsilon} ^{2}\mathcal{K}%
\left( x_{m},x_{m}\right)
\end{equation*}%
Moreover, we have $\mathcal{\tilde{K}}\left( x_{m},x_{m}\right) \mathcal{%
\tilde{K}}\left( x_{m},x_{m}\right) ^{-1}=I_{m}$ or:%
\begin{equation}
I_{m}=\mathcal{K}\left( x_{m},x\right) \mathcal{K}\left( x,x_{m}\right)
\mathcal{\tilde{K}}\left( x_{m},x_{m}\right) ^{-1}+\sigma_{\varepsilon} ^{2}\mathcal{K}%
\left( x_{m},x_{m}\right) \mathcal{\tilde{K}}\left( x_{m},x_{m}\right) ^{-1}
\label{eq:appendix-approx3a}
\end{equation}%
In a similar way, we have $\mathcal{\tilde{K}}\left( x_{m},x_{m}\right) ^{-1}%
\mathcal{\tilde{K}}\left( x_{m},x_{m}\right) =I_{m}$ or:%
\begin{equation}
I_{m}=\mathcal{\tilde{K}}\left( x_{m},x_{m}\right) ^{-1}\mathcal{K}\left(
x_{m},x\right) \mathcal{K}\left( x,x_{m}\right) +\sigma_{\varepsilon} ^{2}\mathcal{\tilde{K%
}}\left( x_{m},x_{m}\right) ^{-1}\mathcal{K}\left( x_{m},x_{m}\right)
\label{eq:appendix-approx3b}
\end{equation}

\subsubsection{Approximation of the conditional expectation}

Using Equations (\ref{eq:appendix-approx1b}) and (\ref{eq:appendix-approx2}%
), we have:%
\begin{eqnarray*}
\hat{m}\left( x^{\star }\right)  &=&\mathcal{K}\left( x^{\star },x\right)
\left( \mathcal{K}\left( x,x\right) +\sigma_{\varepsilon} ^{2}I_{n}\right) ^{-1}y \\
&\approx &\mathcal{K}\left( x^{\star },x_{m}\right) \mathcal{K}\left(
x_{m},x_{m}\right) ^{-1}\mathcal{K}\left( x_{m},x\right) \left( \mathcal{K}%
\left( x,x\right) +\sigma_{\varepsilon} ^{2}I_{n}\right) ^{-1}y \\
&\approx &\mathcal{K}\left( x^{\star },x_{m}\right) \mathcal{K}\left(
x_{m},x_{m}\right) ^{-1}\mathcal{K}\left( x_{m},x\right) \cdot  \\
&&\left( \frac{1}{\sigma_{\varepsilon} ^{2}}I_{n}-\frac{1}{\sigma_{\varepsilon} ^{2}}\mathcal{K}\left(
x,x_{m}\right) \mathcal{\tilde{K}}\left( x_{m},x_{m}\right) ^{-1}\mathcal{K}%
\left( x_{m},x\right) \right) y \\
&=&\frac{1}{\sigma_{\varepsilon} ^{2}}\mathcal{K}\left( x^{\star },x_{m}\right) \mathcal{K}%
\left( x_{m},x_{m}\right) ^{-1}\cdot  \\
&&\left( I_{m}-\mathcal{K}\left( x_{m},x\right) \mathcal{K}\left(
x,x_{m}\right) \mathcal{\tilde{K}}\left( x_{m},x_{m}\right) ^{-1}\right)
\mathcal{K}\left( x_{m},x\right) y
\end{eqnarray*}%
Using Equation (\ref{eq:appendix-approx3a}), we deduce that:%
\begin{eqnarray*}
(\ast ) &=&I_{m}-\mathcal{K}\left( x_{m},x\right) \mathcal{K}\left(
x,x_{m}\right) \mathcal{\tilde{K}}\left( x_{m},x_{m}\right) ^{-1} \\
&=&\mathcal{K}\left( x_{m},x\right) \mathcal{K}\left( x,x_{m}\right)
\mathcal{\tilde{K}}\left( x_{m},x_{m}\right) ^{-1}+\sigma_{\varepsilon} ^{2}\mathcal{K}%
\left( x_{m},x_{m}\right) \mathcal{\tilde{K}}\left( x_{m},x_{m}\right) ^{-1}-
\\
&&\mathcal{K}\left( x_{m},x\right) \mathcal{K}\left( x,x_{m}\right) \mathcal{%
\tilde{K}}\left( x_{m},x_{m}\right) ^{-1} \\
&=&\sigma_{\varepsilon} ^{2}\mathcal{K}\left( x_{m},x_{m}\right) \mathcal{\tilde{K}}\left(
x_{m},x_{m}\right) ^{-1}
\end{eqnarray*}%
Finally, we obtain the approximation of $\hat{m}\left( x^{\star }\right) $:%
\begin{eqnarray*}
\hat{m}\left( x^{\star }\right)  &\approx &\frac{1}{\sigma_{\varepsilon} ^{2}}\mathcal{K}\left(
x^{\star },x_{m}\right) \mathcal{K}\left( x_{m},x_{m}\right) ^{-1}\cdot  \\
&&\sigma_{\varepsilon} ^{2}\mathcal{K}\left( x_{m},x_{m}\right) \mathcal{\tilde{K}}\left(
x_{m},x_{m}\right) ^{-1}\mathcal{K}\left( x_{m},x\right) y \\
&=&\mathcal{K}\left( x^{\star },x_{m}\right) \mathcal{\tilde{K}}\left(
x_{m},x_{m}\right) ^{-1}\mathcal{K}\left( x_{m},x\right) y
\end{eqnarray*}

\subsubsection{Approximation of the conditional covariance}

In the previous paragraph, we have shown that:%
\begin{equation*}
\mathcal{K}\left( x^{\star },x\right) \left( \mathcal{K}\left( x,x\right)
+\sigma_{\varepsilon} ^{2}I_{n}\right) ^{-1}\approx \mathcal{K}\left( x^{\star
},x_{m}\right) \mathcal{\tilde{K}}\left( x_{m},x_{m}\right) ^{-1}\mathcal{K}%
\left( x_{m},x\right)
\end{equation*}%
It follows that:%
\begin{eqnarray*}
\mathcal{\hat{K}}\left( x^{\star },x^{\star }\right)  &=&\mathcal{K}\left(
x^{\star },x^{\star }\right) -\mathcal{K}\left( x^{\star },x\right) \left(
\mathcal{K}\left( x,x\right) +\sigma_{\varepsilon} ^{2}I_{n}\right) ^{-1}\mathcal{K}\left(
x,x^{\star }\right)  \\
&\approx &\mathcal{K}\left( x^{\star },x^{\star }\right) -\mathcal{K}\left(
x^{\star },x_{m}\right) \mathcal{\tilde{K}}\left( x_{m},x_{m}\right) ^{-1}%
\mathcal{K}\left( x_{m},x\right) \mathcal{K}\left( x,x^{\star }\right)
\end{eqnarray*}%
We now replace $\mathcal{K}\left( x,x^{\star }\right) $ and $\mathcal{K}%
\left( x^{\star },x^{\star }\right) $ by Equations (\ref%
{eq:appendix-approx1c}) and (\ref{eq:appendix-approx1d}):%
\begin{eqnarray*}
\mathcal{\hat{K}}\left( x^{\star },x^{\star }\right)  &\approx &\mathcal{K}%
\left( x^{\star },x_{m}\right) \mathcal{K}\left( x_{m},x_{m}\right) ^{-1}%
\mathcal{K}\left( x_{m},x^{\star }\right) - \\
&&\mathcal{K}\left( x^{\star },x_{m}\right) \mathcal{\tilde{K}}\left(
x_{m},x_{m}\right) ^{-1}\mathcal{K}\left( x_{m},x\right) \mathcal{K}\left(
x,x_{m}\right) \mathcal{K}\left( x_{m},x_{m}\right) ^{-1}\mathcal{K}\left(
x_{m},x^{\star }\right)  \\
&=&\mathcal{K}\left( x^{\star },x_{m}\right) \left( I_{m}-\mathcal{\tilde{K}}%
\left( x_{m},x_{m}\right) ^{-1}\mathcal{K}\left( x_{m},x\right) \mathcal{K}%
\left( x,x_{m}\right) \right) \cdot  \\
&&\mathcal{K}\left( x_{m},x_{m}\right) ^{-1}\mathcal{K}\left( x_{m},x^{\star
}\right)
\end{eqnarray*}%
Using Equation (\ref{eq:appendix-approx3a}), we notice that:%
\begin{eqnarray*}
(\ast ) &=&I_{m}-\mathcal{\tilde{K}}\left( x_{m},x_{m}\right) ^{-1}\mathcal{K%
}\left( x_{m},x\right) \mathcal{K}\left( x,x_{m}\right)  \\
&=&\mathcal{\tilde{K}}\left( x_{m},x_{m}\right) ^{-1}\mathcal{K}\left(
x_{m},x\right) \mathcal{K}\left( x,x_{m}\right) +\sigma_{\varepsilon} ^{2}\mathcal{\tilde{K%
}}\left( x_{m},x_{m}\right) ^{-1}\mathcal{K}\left( x_{m},x_{m}\right) - \\
&&\mathcal{\tilde{K}}\left( x_{m},x_{m}\right) ^{-1}\mathcal{K}\left(
x_{m},x\right) \mathcal{K}\left( x,x_{m}\right)  \\
&=&\sigma_{\varepsilon} ^{2}\mathcal{\tilde{K}}\left( x_{m},x_{m}\right) ^{-1}\mathcal{K}%
\left( x_{m},x_{m}\right)
\end{eqnarray*}%
Finally, we obtain:%
\begin{eqnarray*}
\mathcal{\hat{K}}\left( x^{\star },x^{\star }\right)  &\approx &\mathcal{K}\left(
x^{\star },x_{m}\right) \sigma_{\varepsilon} ^{2}\mathcal{\tilde{K}}\left(
x_{m},x_{m}\right) ^{-1}\mathcal{K}\left( x_{m},x_{m}\right) \mathcal{K}%
\left( x_{m},x_{m}\right) ^{-1}\mathcal{K}\left( x_{m},x^{\star }\right)  \\
&=&\sigma_{\varepsilon} ^{2}\mathcal{K}\left( x^{\star },x_{m}\right) \mathcal{\tilde{K}}%
\left( x_{m},x_{m}\right) ^{-1}\mathcal{K}\left( x_{m},x^{\star }\right)
\end{eqnarray*}

\subsection{Hybrid Monte Carlo method}
\label{appendix:hybrid-monte-carlo}

The hybrid Monte Carlo (HMC) algorithm is a special case of Markov Chain Monte
Carlo (MCMC) methods. The objective is to perform sampling from a probability
distribution for which the density and the gradients are known. This approach
was first introduced by Duane \textsl{et al.} (1987) and is also known as
Hamiltonian Monte Carlo (Neal, 2011), since it generates Markov chain states
through Hamiltonian evolution in phase space. More precisely, we note $\pi
\left( q\right) $ a probability distribution over $\mathbb{R}^{n}$. In the
formalism of Hamiltonian mechanics, the state of a system is described through
a position variable $q$, a momentum (velocity) variable $p$:
\begin{equation*}
p=m\frac{\mathrm{d}q}{\mathrm{d}t}
\end{equation*}%
and a function of the two variables, which is called the Hamiltonian:
\begin{equation*}
\mathcal{H}\left( q,p\right) =V\left( q\right) +K\left( p\right)
\end{equation*}%
where $V$ and $K$ are respectively potential and kinetic energies. $K$ is
usually given by $K\left( p\right) =\frac{1}{2}\sum_{i=1}^{n}p_{i}^{2}$. The
Hamiltonian describes entirely the dynamical evolution of the physical system
with Hamilton's equations:
\begin{eqnarray*}
\frac{\mathrm{d}q}{\mathrm{d}t} &=&\frac{\partial \,\mathcal{H}}{\partial \,p} \\
\frac{\mathrm{d}p}{\mathrm{d}t} &=&-\frac{\partial \,\mathcal{H}}{\partial \,q}
\end{eqnarray*}%
Therefore, the evolution of the system is described by the \textit{phase space}
$\left( q,p\right) $. When the time increases, the Hamiltonian remains constant
and the volume is preserved in phase space. Neal (2011) defines the probability
distribution over phase space as:
\begin{equation*}
\mathbb{P}\left( q,p\right) =\frac{1}{C}\exp \left( -\mathcal{H}\left(
q,p\right) \right)
\end{equation*}%
where $C$ is a normalization constant and chooses $V\left( q\right) =-\log
\pi \left( q\right) $. It follows that:%
\begin{equation*}
\left\{
\begin{array}{l}
q\sim \pi  \\
p\sim \mathcal{N}\left( 0,I\right)
\end{array}%
\right.
\end{equation*}%
The idea behind HMC is then simple. We build a mountain that is high for small
values of $\pi $ (\textit{i.e.} places where we would not want to sample often),
and deep for large values of $\pi $, and kick a ball in a random direction. It
will roll on the surface, and is attracted to low values of potential and we
stop it after a fixed time period. Then, we repeat the process, and the
successive positions taken by the ball form the sample of $\pi $.\smallskip

To simulate Hamiltonian evolution, the leap-frog method is usually used. It is
based on Euler's method and finite differences, but is able to enforce
Hamiltonian and volume conservation (Neal, 2011):
\begin{eqnarray*}
p\left( t+\frac{\varepsilon }{2}\right)  &=&p\left( t\right) -\frac{%
\varepsilon }{2}\frac{\partial \,V}{\partial \,q}\left( q\left( t\right)
\right)  \\
q\left( t+\varepsilon \right)  &=&q\left( t\right) +\varepsilon p\left( t+%
\frac{\varepsilon }{2}\right)  \\
p\left( t+\varepsilon \right)  &=&p\left( t+\frac{\varepsilon }{2}\right) -%
\frac{\varepsilon }{2}\frac{\partial \,V}{\partial \,q}\left( q\left(
t+\varepsilon \right) \right)
\end{eqnarray*}%
where the parameters are $\varepsilon >0$  and the number of iterations before
stopping the Hamiltonian dynamics. Once the dynamics is stopped, a new state
$\left( q^{\prime },p^{\prime }\right) $ is proposed. To compensate for
numerical errors in the leap-frog integration, a Metropolis step is carried to
accept the proposed state with probability:
\begin{equation*}
\min \left( 1,\frac{\exp \left( -\mathcal{H}\left( q^{\prime },p^{\prime
}\right) \right) }{\exp \left( -\mathcal{H}\left( q,p\right) \right) }%
\right)
\end{equation*}%
otherwise, the state remains unchanged.

\begin{remark}
In the case of hyperparameter posterior sampling, the distribution $\pi \left(
\theta \right) $ is given by:
\begin{equation*}
\pi \left( \theta \right) =p\left( \theta \mid y\right) =\frac{p\left( y\mid
\theta \right) p\left( \theta \right) }{\int p\left( y\mid \theta ^{\prime
},z\right) p\left( \theta ^{\prime }\right) \,\mathrm{d}\theta ^{\prime }}
\end{equation*}%
where $p\left( \theta \right) $ is the prior distribution on hyperparameters.
Note that the normalization constant is not needed to sample from the posterior
with HMC.
\end{remark}

\subsection{Computation of $\mathbb{E}\left[ \left( X-c\right) ^{+}\right] $}
\label{appendix:black-formula}

We assume that $X\sim \mathcal{N}\left( \mu ,\sigma ^{2}\right) $ and we would
like to calculate $\mathbb{E}\left[ \left( X-c\right) ^{+}\right] $.
We have%
\begin{eqnarray*}
\mathbb{E}\left[ \left( X-c\right) ^{+}\right]  &=&\int_{-\infty }^{+\infty
}\left( x-c\right) \mathds{1}\left\{ x-c\geq 0\right\} \phi \left( x;\mu
,\sigma ^{2}\right) \,\mathrm{d}x \\
&=&\int_{c}^{+\infty }\left( x-c\right) \phi \left( x;\mu ,\sigma
^{2}\right) \,\mathrm{d}x \\
&=&\int_{c}^{+\infty }x\phi \left( x;\mu ,\sigma ^{2}\right) \,\mathrm{d}%
x-c\int_{c}^{+\infty }\phi \left( x;\mu ,\sigma ^{2}\right) \,\mathrm{d}x \\
&=&\int_{c}^{+\infty }\frac{x}{\sigma \sqrt{2\pi }}\exp \left( -\frac{\left(
x-\mu \right) ^{2}}{2\sigma ^{2}}\right) \,\mathrm{d}x-c\left( 1-\Phi \left(
\frac{c-\mu }{\sigma }\right) \right)
\end{eqnarray*}%
By considering the change of variable $y=\sigma ^{-1}\left( x-\mu \right) $,
we obtain:%
\begin{eqnarray*}
(\ast ) &=&\int_{c}^{+\infty }\frac{x}{\sigma \sqrt{2\pi }}\exp \left( -%
\frac{\left( x-\mu \right) ^{2}}{2\sigma ^{2}}\right) \,\mathrm{d}x \\
&=&\int_{\sigma ^{-1}\left( c-\mu \right) }^{+\infty }\frac{\mu +\sigma y}{%
\sqrt{2\pi }}\exp \left( -\frac{1}{2}y^{2}\right) \,\mathrm{d}y \\
&=&\mu \left( 1-\Phi \left( \frac{c-\mu }{\sigma }\right) \right) +\frac{%
\sigma }{\sqrt{2\pi }}\left[ -e^{-y^{2}}\right] _{\sigma ^{-1}\left( c-\mu
\right) }^{+\infty } \\
&=&\mu \left( 1-\Phi \left( \frac{c-\mu }{\sigma }\right) \right) +\sigma
\phi \left( \frac{c-\mu }{\sigma }\right)
\end{eqnarray*}%
Finally, we deduce that:%
\begin{eqnarray*}
\mathbb{E}\left[ \left( X-c\right) ^{+}\right]  &=&\left( \mu -c\right)
\left( 1-\Phi \left( \frac{c-\mu }{\sigma }\right) \right) +\sigma \phi
\left( \frac{c-\mu }{\sigma }\right)  \\
&=&\left( \mu -c\right) \Phi \left( \frac{\mu -c}{\sigma }\right) +\sigma
\phi \left( \frac{\mu -c}{\sigma }\right)
\end{eqnarray*}

\begin{remark}
If we are interested in $\mathbb{E}\left[ \left( c-X\right) ^{+}\right] $,
we use the identity:%
\begin{equation*}
X-c=\left( X-c\right) ^{+}-\left( c-X\right) ^{+}
\end{equation*}%
and we find:
\begin{equation*}
\mathbb{E}\left[ \left( c-X\right) ^{+}\right] =\left( c-\mu \right) \Phi
\left( \frac{c-\mu }{\sigma }\right) +\sigma \phi \left( \frac{c-\mu }{%
\sigma }\right)
\end{equation*}
\end{remark}

\subsection{Improvement-based minimization problem}
\label{appendix:improvement-minimization}

If we are interested in finding the minimum, we define the improvement as $%
\Delta _{n}\left( x^{\star }\right) =\left( \tau -\hat{f}_{n}\left( x^{\star
}\right) \right) ^{+}$. It follows that:%
\begin{equation*}
\Pr \left\{ \Delta _{n}\left( x^{\star }\right) >0\right\} =\Phi \left(
\frac{\tau -\hat{m}_{n}\left( x^{\star }\right) }{\sqrt{\mathcal{\hat{K}}%
_{n}\left( x^{\star },x^{\star }\right) }}\right)
\end{equation*}%
and:%
\begin{equation*}
\limfunc{EI}\nolimits_{n}\left( x^{\star }\right) =\left( \tau -\hat{m}%
_{n}\left( x^{\star }\right) \right) \Phi \left( \frac{\tau -\hat{m}%
_{n}\left( x^{\star }\right) }{\sqrt{\mathcal{\hat{K}}_{n}\left( x^{\star
},x^{\star }\right) }}\right) +\sqrt{\mathcal{\hat{K}}_{n}\left( x^{\star
},x^{\star }\right) }\phi \left( \frac{\tau -\hat{m}_{n}\left( x^{\star
}\right) }{\sqrt{\mathcal{\hat{K}}_{n}\left( x^{\star },x^{\star }\right) }}\right)
\end{equation*}%
Therefore, we have $x_{n+1}=\arg \max\, \mathcal{U}_{n}\left( x^{\star }\right)
$ where $\mathcal{U}_{n}\left( x^{\star }\right) =\Pr \left\{ \Delta _{n}\left(
x^{\star }\right) >0\right\} $ or $\mathcal{U}_{n}\left( x^{\star }\right)
=\limfunc{EI}\nolimits_{n}\left( x^{\star }\right) $. The standard problem is
obtained by setting $\tau =f_{n}\left( \varkappa _{n}^{\star }\right) $ where
$\varkappa _{n}^{\star }=\arg \min_{\varkappa \in x}f\left( \varkappa \right)$.

\subsection{Matrix-variate Student-$t$ distribution}
\label{appendix:mvt}

We first recall that a random vector $X\in \mathbb{R}^{n}$ has a multivariate
Student-$t$ distribution with mean $\mu $ and scale matrix $\Sigma $ if its
probability density function is equal to:
\begin{equation*}
f\left( x\right) =\frac{\Gamma \left( \upsilon /2\right) }{\left( \nu \pi
\right) ^{n/2}\Gamma \left( \left( \upsilon -n\right) /2\right) }\left\vert
\Sigma \right\vert ^{-1/2}\left( 1+\frac{1}{\nu }y^{\top }\Sigma
^{-1}y\right) ^{-\upsilon /2}
\end{equation*}%
where $y=x-\mu $, $\upsilon =\nu +n$ and $\nu $ is the degrees of freedom. Let
$X$ be an $n\times p$ random matrix. It has a matrix-variate Student-$t$
distribution with mean matrix $M\in \mathbb{R}^{n\times p}$ and covariance
matrices $\Sigma \in \mathbb{R}^{n\times n}$ and $\Omega \in \mathbb{R}
^{p\times p}$ if its probability density function is equal to (Gupta and Nagar,
1999):
\begin{eqnarray*}
f\left( X\right)  &=&\frac{\Gamma _{n}\left( \upsilon /2\right) }{\left( \nu
\pi \right) ^{np/2}\Gamma _{n}\left( \left( \upsilon -p\right) /2\right) }%
\left\vert \Sigma \right\vert ^{-p/2}\left\vert \Omega \right\vert
^{-n/2}\cdot  \\
&&\left\vert I_{n}+\Sigma ^{-1}\left( X-M\right) \Omega ^{-1}\left(
X-M\right) \right\vert ^{-\upsilon /2}
\end{eqnarray*}%
where:%
\begin{equation*}
\upsilon =\nu +n+p-1
\end{equation*}%
We note $X\sim \mathcal{MT}_{n,p}\left( M,\Sigma ,\Omega ;\nu \right) $. This
matrix distribution has several properties similar to Gaussian random matrices,
which make it suitable for multivariate regression. For example, we have:
\begin{equation*}
\mathbb{E}\left[ X\right] =M
\end{equation*}%
and:%
\begin{equation*}
\limfunc{cov}\left( \limfunc{vec}X^{\top }\right) =\frac{1}{\nu -2}\Sigma
\otimes \Omega \text{\qquad if }v>2
\end{equation*}%
Gupta and Nagar (1999) also show that:
\begin{equation*}
X^{\top }\sim \mathcal{MT}_{p,n}\left( M^{\top },\Sigma ,\Omega ;\nu \right)
\end{equation*}%
If we assume that:%
\begin{equation*}
\left(
\begin{array}{c}
X \\
Y%
\end{array}%
\right) \sim \mathcal{MT}_{n,p}\left( \left(
\begin{array}{c}
M_{x} \\
M_{y}%
\end{array}%
\right) ,\left(
\begin{array}{cc}
\Sigma _{xx} & \Sigma _{xy} \\
\Sigma _{yx} & \Sigma _{yy}%
\end{array}%
\right) ,\Omega ;\nu \right)
\end{equation*}%
we have:%
\begin{equation*}
X\sim \mathcal{MT}_{n_{x},p}\left( M_{x},\Sigma _{xx},\Omega ;\nu \right)
\end{equation*}%
and:%
\begin{equation*}
Y\sim \mathcal{MT}_{n_{y},p}\left( M_{y},\Sigma _{yy},\Omega ;\nu \right)
\end{equation*}%
Moreover, the conditional distribution is still a matrix-variate Student-$t$
distribution:
\begin{equation*}
Y\mid X=x\sim \mathcal{MT}_{n_{y},p}\left( M_{y\mid x},\Sigma _{yy\mid
x},\Omega _{yy\mid x};\nu +n_{x}\right)
\end{equation*}%
where:%
\begin{equation*}
\left\{
\begin{array}{l}
M_{y\mid x}=M_{y}+\Sigma _{yx}\Sigma _{xx}^{-1}\left( x-M_{x}\right)  \\
\Sigma _{yy\mid x}=\Sigma _{yy}-\Sigma _{yx}\Sigma _{xx}^{-1}\Sigma _{xy} \\
\Omega _{yy\mid x}=\Omega +\left( x-M_{x}\right) ^{\top }\Sigma
_{xx}^{-1}\left( x-M_{x}\right)
\end{array}%
\right.
\end{equation*}%
If we consider a column-based partition:%
\begin{equation*}
\left(
\begin{array}{cc}
X & Y%
\end{array}%
\right) \sim \mathcal{MT}_{n,p}\left( \left(
\begin{array}{cc}
M_{x} & M_{y}%
\end{array}%
\right) ,\Sigma ,\left(
\begin{array}{cc}
\Omega _{xx} & \Omega _{xy} \\
\Omega _{yx} & \Omega _{yy}%
\end{array}%
\right) ;\nu \right)
\end{equation*}%
we have:%
\begin{equation*}
X\sim \mathcal{MT}_{n,p_{x}}\left( M_{x},\Sigma ,\Omega _{xx};\nu \right)
\end{equation*}%
and:%
\begin{equation*}
Y\sim \mathcal{MT}_{n,p_{y}}\left( M_{y},\Sigma ,\Omega _{yy};\nu \right)
\end{equation*}%
For the conditional distribution, Gupta and Nagar (1999) show that:%
\begin{equation*}
Y\mid X=x\sim \mathcal{MT}_{n,p_{y}}\left( M_{y\mid x},\Sigma _{yy\mid
x},\Omega _{yy\mid x};\nu +p_{x}\right)
\end{equation*}%
where:%
\begin{equation*}
\left\{
\begin{array}{l}
M_{y\mid x}=M_{y}+\left( x-M_{x}\right) \Omega _{xx}^{-1}\Omega _{xy} \\
\Sigma _{yy\mid x}=\Sigma +\left( x-M_{x}\right) \Omega _{xx}^{-1}\left(
x-M_{x}\right) ^{\top } \\
\Omega _{yy\mid x}=\Omega _{yy}-\Omega _{yx}\Omega _{xx}^{-1}\Omega _{xy}%
\end{array}%
\right.
\end{equation*}

\subsection{ADMM algorithm}
\label{appendix:admm}

The alternating direction method of multipliers (ADMM) is an algorithm
introduced by Gabay and Mercier (1976) to solve problems which can be expressed
as\footnote{We follow the standard presentation of Boyd \textsl{et al.} (2011)
on ADMM.}:
\begin{eqnarray}
\left\{ x^{\star },z^{\star }\right\}  &=&\arg \min f\left( x\right)
+g\left( z\right)   \label{eq:appendix-admm1} \\
&\text{s.t.}&Ax+Bz-c=0  \notag
\end{eqnarray}%
where $A\in \mathbb{R}^{p\times n}$, $B\in \mathbb{R}^{p\times m}$, $c\in
\mathbb{R}^{p}$, and the functions $f:\mathbb{R}^{n}\rightarrow \mathbb{R}%
\cup \{+\infty \}$ and $g:\mathbb{R}^{m}\rightarrow \mathbb{R}\cup \{+\infty
\}$ are proper closed convex functions. Boyd \textsl{et al.} (2011) show that
the ADMM algorithm consists of three steps:

\begin{enumerate}
\item The $x$-update is:
\begin{equation}
x^{\left( k\right) }=\arg \min \left\{ f\left( x\right) +\frac{\varphi }{2}%
\left\Vert Ax+Bz^{\left( k-1\right) }-c+u^{\left( k-1\right) }\right\Vert
_{2}^{2}\right\}   \label{eq:appendix-admm2a}
\end{equation}

\item The $z$-update is:%
\begin{equation}
z^{\left( k\right) }=\arg \min \left\{ g\left( z\right) +\frac{\varphi }{2}%
\left\Vert Ax^{\left( k\right) }+Bz-c+u^{\left( k-1\right) }\right\Vert
_{2}^{2}\right\}   \label{eq:appendix-admm2b}
\end{equation}

\item The $u$-update is:%
\begin{equation}
u^{\left( k\right) }=u^{\left( k-1\right) }+\left( Ax^{\left( k\right)
}+Bz^{\left( k\right) }-c\right)   \label{eq:appendix-admm2c}
\end{equation}
\end{enumerate}
In this approach, $u^{\left( k\right) }$ is the dual variable of the primal
residual $r=Ax+Bz-c$ and $\varphi $ is the $\boldsymbol{\ell}_{2}$ penalty
variable. In the paper, we use the notations $f^{\left( k\right) }\left(
x\right) $ and $g^{\left( k\right) }\left( z\right) $ when referring to the
objective functions that are defined in the $x$- and $z$-steps.

\section{Software}

Throughout this article, we used of the following open-source software
libraries:
\begin{itemize}
\item GPML Matlab Toolbox (Rasmussen and Nickisch, 2010)

\item GPy: A Gaussian process framework in Python
\begin{center}
\url{http://github.com/SheffieldML/GPy}
\end{center}

\item GPyOpt: A Bayesian Optimization Framework in Python
\begin{center}
\url{http://github.com/SheffieldML/GPyOpt}
\end{center}

\end{itemize}

\section{Additional figures}

\begin{figure}[tbph]
\centering
\caption{Estimated ridge penalization $\lambda _{t}$ (in \%)}
\label{fig:trend-lambda}
\figureskip
\includegraphics[width = \figurewidth, height = \figureheight]{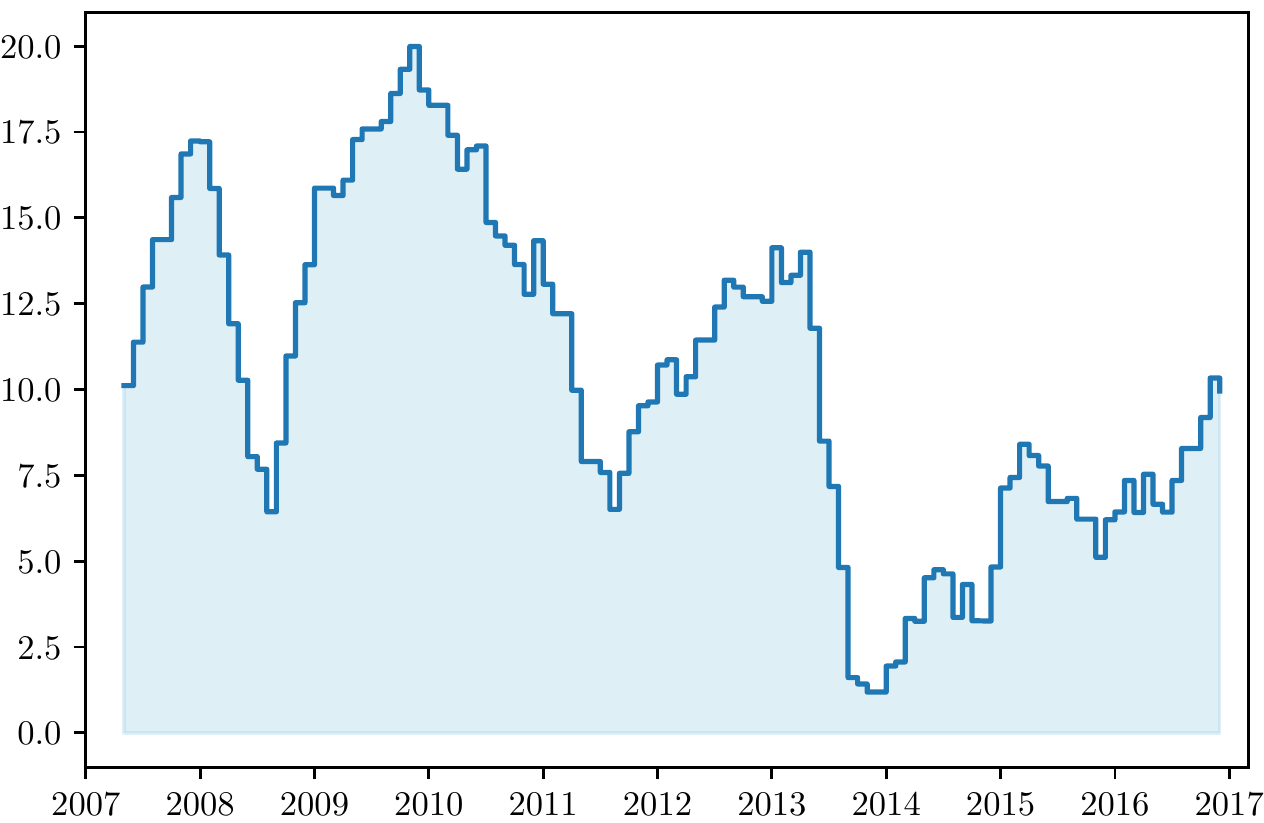}
\end{figure}

\begin{figure}[tbph]
\centering
\caption{Estimated return window length $\ell _{t}\left( \mu \right) $ (in months)}
\label{fig:trend-mu}
\figureskip
\includegraphics[width = \figurewidth, height = \figureheight]{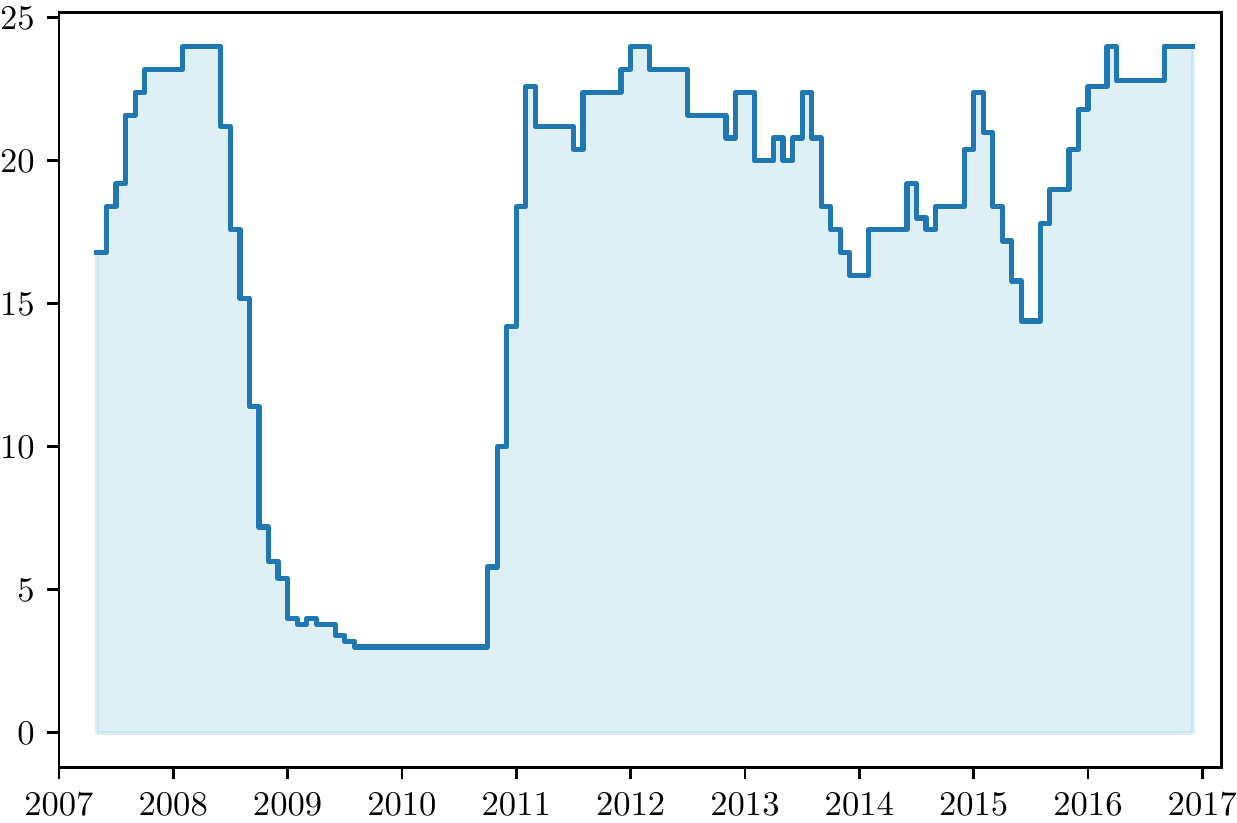}
\end{figure}

\begin{figure}[tbph]
\centering
\caption{Estimated covariance window length $\ell _{t}\left( \Sigma \right) $ (in months)}
\label{fig:trend-cov}
\figureskip
\includegraphics[width = \figurewidth, height = \figureheight]{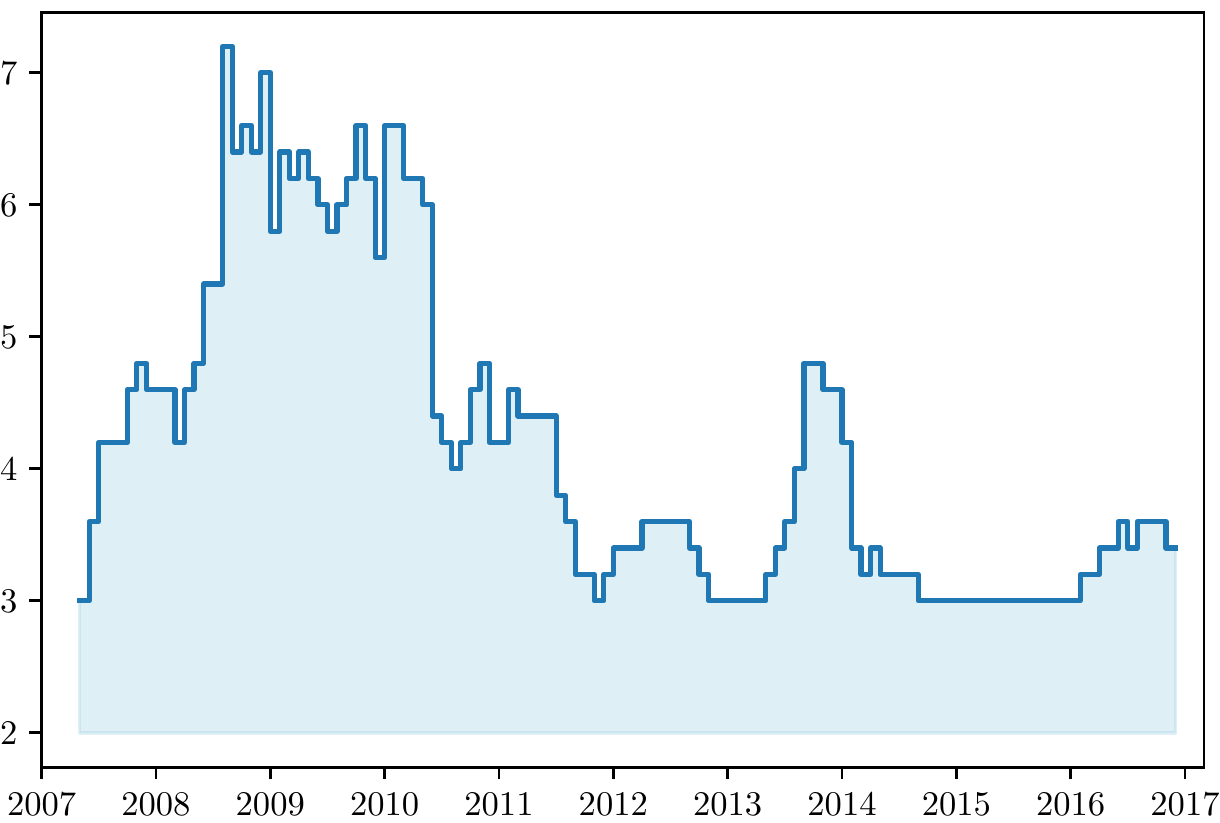}
\end{figure}

\end{document}